\documentclass[aps,prd,10pt,superscriptaddress,onecolumn,nolongbibliography,shortbibliography,showkeys,serif]{revtex4-2}

\usepackage{graphicx}
\usepackage{xcolor}
\usepackage{hyperref}
\usepackage{amsfonts, amsmath, amssymb}
\usepackage{lmodern}
\usepackage{dcolumn}
\usepackage{multirow}
\usepackage{enumitem}
\usepackage{booktabs}
\usepackage{enumitem}
\usepackage{subcaption}
\usepackage[sort&compress]{natbib}

\renewcommand{\d}{\mathrm{d}}
\newcommand{\orcid}[1]{\href{https://orcid.org/#1}{\includegraphics[width=8pt]{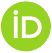}}}

\hypersetup{colorlinks=true, linkcolor=blue, urlcolor=blue, citecolor=blue}

\graphicspath{{figs/}} 

\begin{document}

\title{New Expansion Rate Anomalies at Characteristic Redshifts Geometrically Determined using DESI-DR2 BAO and DES-SN5YR Observations}

\author{Purba Mukherjee \orcid{0000-0002-2701-5654}}
\email{pdf.pmukherjee@jmi.ac.in}
\affiliation{Centre for Theoretical Physics, Jamia Millia Islamia, New Delhi-110025, India}%

\author{Anjan A Sen \orcid{0000-0001-9615-4909}}%
\email{aasen@jmi.ac.in}
\affiliation{Centre for Theoretical Physics, Jamia Millia Islamia, New Delhi-110025, India}%

%\date{\today}% 

\begin{abstract}
We perform a model-independent reconstruction of the  cosmic distances using the Multi-Task Gaussian Process (MTGP) framework  as well as knot-based spline techniques with DESI-DR2 BAO and DES-SN5YR datasets. We calibrate the comoving sound horizon at the baryon drag epoch $r_d$ to the Planck value, ensuring consistency with early-universe physics. With the reconstructed cosmic distances and their derivatives, we obtain  seven characteristic redshifts in the range $0.3 \leq z \leq 1.7$. We derive the normalized expansion rate of the Universe $E(z)$ at these redshifts. Our findings reveal a significant deviations of approximately $4$ to 5$\sigma$ from the Planck 2018 $\Lambda$CDM predictions,  particularly pronounced in the redshift range $z \sim 0.35$–0.55. These anomalies are consistently observed across both reconstruction methods and combined datasets, indicating robust late-time tensions in the expansion rate of the Universe and which are distinct from the existing ``Hubble Tension''. This could signal new physics beyond the standard cosmological framework at this redshift range.  Our findings underscore the role of characteristic redshifts as sensitive indicators of expansion rate anomalies and motivate further scrutiny with forthcoming datasets from DESI-5YR BAO, Euclid, and LSST. These future surveys will tighten constraints and will confirm whether these late-time anomalies arise from new fundamental physics or unresolved systematics in the data.
\end{abstract}

\keywords{cosmology -- dark energy -- cosmological constant -- cosmological parameters}

\maketitle

%\tableofcontents

\section{Introduction \label{sec:intro}}

In recent decades, cosmology has witnessed remarkable progress, largely driven by high-precision observations from the Cosmic Microwave Background (CMB) \cite{Planck:2018vyg, ACT:2020gnv, Tristram:2023haj}, type Ia supernovae (SNIa) \cite{Blanchard:2022xkk, Peebles:2024txt}, large-scale structure (LSS), and galaxy surveys \cite{BOSS:2014hhw, BOSS:2016wmc, eBOSS:2020yzd, Addison:2017fdm}. The standard model of cosmology, $\Lambda$CDM, describes the universe as being predominantly composed of cold dark matter (CDM) and dark energy (DE), with the latter represented by a cosmological constant ($\Lambda$) that drives the late-time accelerated expansion of the universe \cite{Weinberg:1988cp, Sahni:1999gb}.

Despite its successes, the $\Lambda$CDM model has faced several challenges, as recent observations have revealed notable inconsistencies \cite{Hazra:2013dsx, Verde:2019ivm, Bull:2015stt, Perivolaropoulos:2021jda, Brieden:2022heh, Efstathiou:2024dvn}. One of the most debated issues is the Hubble tension \cite{DiValentino:2020zio, Riess:2021jrx}, where local measurements of the Hubble constant ($H_0$) differ by more than $5\sigma$ from values inferred using CMB data. Another discrepancy, known as the $\sigma_8$ tension \cite{DiValentino:2020vvd}, refers to a $\approx 2.5\sigma$ mismatch between the CMB-predicted vs observed clustering of matter, measured by galaxy surveys \cite{DES:2021wwk,Heymans:2020gsg,Li:2023tui}. Furthermore, deep-space observations from the James Webb Space Telescope (JWST) have uncovered unexpectedly massive and bright galaxies at redshifts $z \gtrsim 7$ \cite{Boylan-Kolchin:2022kae, Labbe:2022ahb, Sabti:2023xwo}, posing additional challenges to the concordance framework. These discrepancies indicate potential gaps in our understanding of cosmic evolution and structure formation, suggesting the need for extensions or modifications beyond the conventional six-parameter $\Lambda$CDM model \cite{DiValentino:2021izs, Abdalla:2022yfr}.  

The latest challenge for the $\Lambda$CDM model has emerged from the highly precise redshift measurements of millions of galaxies by the Dark Energy Spectroscopic Instrument (DESI) with extraordinary accuracy. These measurements enable both isotropic and anisotropic determinations of the Baryon Acoustic Oscillation (BAO) scale across multiple redshift bins, covering the range $0.1 < z < 3.5$  \cite{DESI:2025zpo}.  The recent DESI-DR2 BAO data, when combined with CMB measurements by Planck-2018 as well as various SN-Ia measurements (Pantheon-Plus \cite{Brout:2022vxf}/ DES-SN5YR \cite{DES:2024jxu}/ Union3 \cite{Rubin:2023ovl}), give hints of dynamical dark energy at $\sim 2.8 - 4.2\sigma$ level, putting a serious question mark on the validity of $\Lambda$CDM \cite{DESI:2025zgx}. Although the primary result for the evidence dynamical dark energy comes when one uses the well-studied Chevallier-Polarski-Linder (CPL) parametrization \cite{Chevallier:2000qy} to model the dark energy equation of state (EoS), there have also been model-independent analysis that show evidence of dynamical dark energy \cite{DESI:2025fii, DESI:2025wyn, Dinda:2025svh, You:2025uon, Ormondroyd:2025iaf, Nesseris:2025lke, Berti:2025phi,Yang:2025mws, Gao:2025ozb, Gomez-Valent:2023uof, Yang:2025kgc}. 

Conversely, the recent ACT measurements of CMB fluctuations, especially at small scales, have confirmed that there is no conclusive evidence for any new physics (compared to Planck-2018 results) at the decoupling epoch (during the formation of CMB) \cite{ACT:2025fju, ACT:2025tim}. Consequently, while looking for a solution for Hubble Tension, or the anomalous JWST observations, or the low-redshift measurements for background evolution by DESI, DES, Union3, Pantheon-Plus, and others, the focus now shifts more toward possible modifications of the $\Lambda$CDM model at late times \cite{Scherer:2025esj, Ye:2025oms, Pan:2025qwy, Chen:2025wwn, Shah:2025ayl, Choudhury:2025bnx, Shah:2024gfu, Abedin:2025dis, Postolak:2025qmv, Cheng:2025lod, Wang:2025zri, Specogna:2025guo, RoyChoudhury:2025dhe, Wolf:2025jed, Duchaniya:2025oeh, Colgain:2025nzf, Kessler:2025kju, Chakraborty:2025syu, Chavan:2025xcq, Hossain:2025grx, Bansal:2025ipo, Sakr:2025fay, Ferrari:2025egk, Murai:2025msx, Tyagi:2025zov, Borghetto:2025jrk, Richarte:2025tkj, Liu:2025mub, Richarte:2025tkj, VanRaamsdonk:2025wvj, Goswami:2025uih, Park:2025azv, Keeley:2025stf, Dinda:2025iaq, Yang:2025jei, Giani:2025hhs, Akrami:2025zlb, Gomez-Valent:2024ejh, Sen:2021wld, Mukherjee:2025myk}. 

At the same time, caution must be exercised when combining different observational datasets to search for evidence of deviations from the $\Lambda$CDM model. Potential tensions between datasets, if truly genuine, may limit the validity of combining them for drawing robust conclusions about new physics. Likewise, systematic uncertainties within individual datasets can compromise the reliability of cosmological inferences if not properly accounted for. So, while studying evidence for new physics, we need to be careful when combining data from different observational setups to ensure their consistency \cite{Afroz:2025iwo, Yang:2025qdg, Teixeira:2025czm, Wang:2025bkk, Alfano:2025gie, Keil:2025ysb, Park:2024vrw, Bousis:2024rnb, Ye:2025ark, DESI:2025qqy, DES:2024ywx, Bousis:2024rnb, Cortes:2025joz, Singh:2025seo, Efstathiou:2025tie, Efstathiou:2024xcq, Colgain:2024mtg, Colgain:2024ksa, Ormondroyd:2025exu, DES:2025tir, Sapone:2024ltl, Pedrotti:2024kpn, DESI:2024ude, Huang:2024qno, Gao:2024ily, RoyChoudhury:2024wri, Li:2024qso, Paliathanasis:2025dcr, Paliathanasis:2025hjw, Luciano:2025hjn}.

Although there are multiple studies in the literature, as stated above, to look for evidence for beyond the $\Lambda$CDM model---either by employing various parameterizations of dark energy or through model-independent, nonparametric approaches---it is always interesting to look for new tensions/anomalies in the observational data or cosmological models, specially at low redshifts (where most of the high-precision recent cosmological observations are available), which has not been studied as yet \cite{CosmoVerse:2025txj}. This is particularly interesting as we now have a large collection of very accurate cosmological observations at low redshifts that allow us to measure cosmological observables like co-moving distances, angular diameter distances or luminosity distances very precisely. This helps us to study in a model-independent way any tensions/anomalies present in these measurements at some particular redshifts compared to the prediction from the flat $\Lambda$CDM model as constrained by Planck-2018. 

In this paper,   we use the geometric observables, e.g luminosity distance $D_L(z)$, angular diameter distance $D_A(z)$ and comoving distance $D_M(z)$---to reconstruct the normalized expansion rate $E(z)$ and angular diameter distance $D_A(z)$ across the redshift range  $0.3 \leq z \ 1.7$. We employ two complementary methods: multitask Gaussian processes \cite{Haridasu:2018gqm, Perenon:2021uom, Mukherjee:2024ryz, Mukherjee:2024pcg, Dinda:2024ktd} and  flexible spline-based knot interpolation \cite{Jiang:2024xnu, Ormondroyd:2025exu, Ormondroyd:2025iaf, Bansal:2025ipo, Berti:2025phi}, to reconstruct the cosmic distances and their derivatives. These allow us to identify seven $7$ characteristic redshifts, where model-independent constraints from data are particularly robust and examine how these compare to the expectations from Planck $\Lambda$CDM.  Additionally, we perform several cross-checks to ensure the robustness of our findings. These include consistency tests within the DESI-DR2 BAO dataset, checks for violations of the distance duality relation (CDDR), and an investigation into whether the observed features can be attributed to systematics in the data. Thus, our approach acts as a robust, geometry-based probe to investigate potential hints for departures from the standard concordance framework.

This paper is organized as follows: In Sec. \ref{sec:distances}, we introduce the concept of characteristic redshifts derived from cosmic distance-derivatives. Sec. \ref{sec:method} outlines the reconstruction methodology employed in this work, followed by a summary of the observational datasets used in Sec. \ref{sec:data}. In Sec. \ref{sec:recon}, we present our reconstruction of the cosmic distance-derivatives and related cosmological quantities. Sec. \ref{sec:analysis} discusses potential signs of new physics inferred from our model-independent analysis. In Sec. \ref{sec:anomalies}, we examine possible anomalies in the DESI DR2 BAO measurements. Sec. \ref{sec:cddr} revisits the cosmic distance-duality relation in light of our reconstructions. Finally, we summarize our main findings and make some concluding remarks in Sec. \ref{sec:summary}.

\section{Characteristic Redshifts from Cosmic Distance-Derivatives  \label{sec:distances}}

The universe on a large scale is described by the spatially flat, homogeneous and isotropic Friedmann-Lema\^{i}tre-Robertson-Walker (FLRW) metric, whose line element in comoving coordinates is given by
\begin{equation}
    \d s^2 = -\d t^2 + a^2(t) \, \left[ \, \d r^2 + r^2 \, \d \Omega^2 \right] \, ,
\end{equation}
where $a(t)$ is the scale factor for the expanding Universe $r$ is the comoving radial coordinate, and $\d\Omega^2 = \d\theta^2 + \sin^2\theta\, \d\phi^2$ is the metric on the unit sphere. In an ever-expanding Universe, $a(t)$ is a monotonic function of time and can be related to the cosmological redshift as $1+z = a^{-1}(t)$ where we have normalised the present value of the scale factor $a(t_0) = a_0=1$  (in rest of the paper, quantities with subscript `0' will denote values at present).

In this FLRW setting, the physical distance between two infinitesimally separated events at the same cosmic time ($\d t = 0$) is given by $\d s_p = a(t)\, \d r$. However, in cosmology, direct measurements of physical distances are not possible. Instead, we infer distances through observables such as redshift, angular size, and flux, which leads to defining three key distance measures \cite{Hogg:1999ad}, namely the comoving distance, angular diameter distance and luminosity distance. 

For a light ray traveling radially ($\d \Omega = 0$), the condition for a null geodesic ($\d s^2 = 0$) gives
\begin{equation}
    {\d r} = \frac{c \,\d t}{a(t)} \, .
\end{equation}
Integrating this along the path of the photon, the comoving distance to an object emitting light at time $t$ (or equivalently redshift $z$) is defined as
\begin{equation}
    d_M(z) \equiv c \int_{t}^{t_0} \frac{\d t}{a(t)} = c \int_0^z \frac{\d z'}{H(z')},
\end{equation}

The angular diameter distance $d_A$ is the ratio of an object's transverse size to its observed angular size. It is related to $d_M$ as 
\begin{equation}
    d_A(z) = \frac{d_M(z)}{1+z} \, .
\end{equation}
The luminosity distance $d_L$ is defined through the flux–luminosity relation, accounting for redshift effects on energy and photon arrival rate. It is related to $d_M$ as 
\begin{equation}
    d_L(z) = {(1+z)} \, {d_M(z)} \, .
\end{equation}

\begin{figure}
    \centering
    \includegraphics[width=0.7\linewidth]{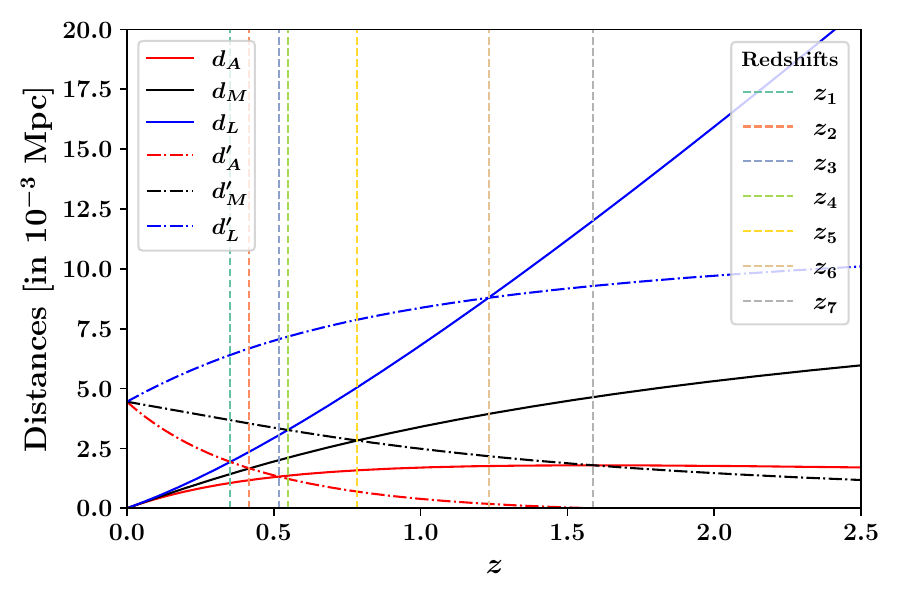}
    \caption{Evolution of the cosmic distances: $d_A(z)$, $d_M(z)$ and $d_L(z)$, and their derivatives ($d_A'(z)$, $d_M'(z)$ and $d_L'(z)$), for the Planck 2018 baseline $\Lambda$CDM model. Here, $z_i$ ($i = 1 \cdots 7$) are the characteristic redshifts.}
    \label{fig:lcdm}
\end{figure}

Therefore, in FLRW geometry, cosmic distances are determined entirely by the expansion history encapsulated in the Hubble parameter $H(z)$ (more fundamentally in the scale factor $a(t)$). Beyond the definitions of distances themselves, their derivatives with respect to (\textit{w.r.t}) redshift contain further geometric information. On assuming the Planck 2018 baseline $\Lambda$CDM model \cite{Planck:2018vyg}, we can plot the evolution of $d_A(z)$, $d_M(z)$, and $d_L(z)$, as well as their derivatives ($d_A'(z)$, $d_M'(z)$ and $d_L'(z)$ where `$\prime$' denotes derivative \textit{w.r.t.} redshift $z$) (shown in Fig. \ref{fig:lcdm}). For a flat universe, the simplicity of the distance relations allows one to identify redshifts at which specific derivative-based identities hold. We define the following redshifts $z_i$ via these conditions:
\begin{align} \label{eq:z_char}
    & d_A'(z_1) = d_L(z_1)  \, , \nonumber\\ % z4
    & d_A'(z_2) = d_M(z_2)  \, , \nonumber\\ % z3
    & d_A'(z_3) = d_A(z_3)  \, , \nonumber\\ % z2
    & d_M'(z_4) = d_L(z_4)  \, , \\ % z7
    & d_M'(z_5) = d_M(z_5)  \, , \nonumber\\ % z5
    & d_L'(z_6) = d_L(z_6)  \, , \nonumber\\ % z8
    & d_M'(z_7) = d_A(z_7) \quad \text{or} \quad d_A'(z_7) = 0 \quad \implies \quad  d_A(z_7) \longrightarrow \text{max.}  \, .  \nonumber % z1 and z6
    \end{align}
We refer to these as \textit{characteristic redshifts} - geometrically defined points in the redshift range where two cosmic distances (or a derivative and a distance) coincide. These identities arise from the mathematical structure of distance definitions in the flat FLRW background and do not rely on any assumptions about the matter content or dark energy model. However, the values of these characteristic redshifts $z_i$ serve as diagnostic indicators that are sensitive to the underlying cosmological model. 

Crucially, at these redshifts, the Hubble parameter can be algebraically isolated from the distance measures and their derivatives. Thus, we can compute $H(z_i)$ at each $z_i$ as functions of $d_A(z_i)$, viz.
\begin{align} \label{eq:Hz_char}
    & H(z_1) = \frac{c}{d_A(z_1) \left[ (1+z_1)^{3} + 1\right]}  \, , \nonumber \\ % z4
    & H(z_2) = \frac{c}{d_A(z_2) \left[ (1+z_1)^{2} + 1\right]}  \, , \nonumber\\ % z3
    & H(z_3) = \frac{c}{d_A(z_3) \left[ 2 + z_3\right]}  \, , \nonumber\\ % z2
    & H(z_4) = \frac{c}{d_A(z_4) \left[ (1+z_4)^{2} \right]}  \, , \\ % z7
    & H(z_5) = \frac{c}{d_A(z_5) \left[ 1+z_1 \right]}  \, , \nonumber\\ % z5
    & H(z_6) = \frac{c}{d_A(z_6) \, z_6}  \, , \nonumber\\ % z8
    & H(z_7) = \frac{c}{d_A(z_7)}  \, . \nonumber % z1 and z6
\end{align}
This provides a pathway for direct inference of the Hubble parameter--- constructed from a more fundamental quantity like the scale factor--- independent of assumptions about the full shape of the expansion history or cosmological parameters. 

Once the Hubble parameter $H(z)$ is identified at each of the characteristic redshifts through distance derivative identities, we can use this structure to extract expansion history information in a model-independent fashion. In this work, we make direct use of the latest DESI-DR2 BAO \cite{DESI:2025zgx} and DES-5YR SNIa \cite{DES:2024jxu} data to reconstruct the comoving distance $d_M(z)$ without assuming any specific cosmological model. We also the early universe prior on angular scale of sound horizon at recombination given by Planck-2018 result \cite{Planck:2018vyg} for flat $\Lambda$CDM model. The details of the data used in our analysis are provided in Section \ref{sec:data}.

We adopt two complementary approaches to this end: (i) a non-parametric reconstruction of $d_M(z)$ using Gaussian processes (GP), which smoothly interpolates the data while accounting for correlated uncertainties \cite{Haridasu:2018gqm, Perenon:2021uom, Mukherjee:2024ryz, Mukherjee:2024pcg, Dinda:2024ktd}, and (ii) a free-form reconstruction where the $d_M(z)$ is specified by interpolating between discrete nodes (or ``knots'') \cite{Jiang:2024xnu, Ormondroyd:2025exu, Ormondroyd:2025iaf, Bansal:2025ipo, Berti:2025phi} placed precisely at the characteristic redshifts discussed earlier. Both approaches allow us to compute derivatives of the distances \textit{w.r.t} redshift in a controlled and data-driven manner, enabling a direct and robust inference of $H(z)$ at these characteristic redshifts of geometric coincidences.

\section{Reconstruction Methodology \label{sec:method}} 

We now turn to outlining the methodology used to reconstruct cosmic distances and their derivatives. To ensure both robustness and flexibility in our analysis, we adopt two complementary reconstruction techniques: Gaussian processes (GP) and a free-form knot-based spline interpolation. These methods are chosen for their ability to model the complex data in a flexible, cosmology-independent manner while consistently accounting for observational uncertainties and propagating them to all derived quantities.
\begin{itemize}[left=0pt]
\item \textbf{Gaussian processes:} GP regression is a Bayesian, non-parametric technique that models functions based on a prior over smooth distributions, informed by the data. For a detailed discussion of the GP framework and its implementation, we refer the reader to Refs. \cite{rw, PhysRevD.84.083501, PhysRevLett.105.241302, Seikel:2012uu, Shafieloo:2012ht}. We implement two variants of GP regression:
\begin{itemize}
    \item \emph{Single-task} setup (STGP): To reconstruct a single target function, denoted $f(x)$, from observed data. This allows for direct modelling of the function and its derivatives. Uncertainties from the input data are automatically propagated to the prediction and its derivative through the GP framework.

    \item \emph{Multi-task} setup (MTGP): To reconstruct multiple correlated functions, such as $f(x)$ and its derivative $f'(x)$, or two related observables, by placing a joint prior over them. This allows for joint modelling with correlated uncertainties, improving inference when multiple datasets or data modalities are available.
\end{itemize}
Several covariance kernels are tested, including the Matérn-$\nu$ family with $\nu = 7/2$ (M72) and $\nu = 9/2$ (M92),
\begin{equation}
    \kappa(x, \tilde{x}) = \sigma_f^2 \frac{\left[ \frac{\sqrt{2 \nu}}{l} (x - \tilde{x})\right]^\nu}{\mathrm{\Gamma}(\nu)\, 2^{\nu -1}} \mathrm{K}_{\nu} \left( \frac{\sqrt{2\nu}}{l}(x-\tilde{x})\right) \, ,
\end{equation}
the squared exponential (RBF),
\begin{equation}
    \kappa(x, \tilde{x}) = \sigma_f^2 \exp \left[ - \frac{(x - \tilde{x})^2}{2l^2}\right] \, ,
\end{equation}
and the rational quadratic (RQ) kernel,
\begin{equation}
    \kappa(x, \tilde{x}) = \sigma_f^2 \left[ 1 + \frac{(x - \tilde{x})^2}{2\alpha l^2}\right]^{-\alpha} \, .
\end{equation}
Note, $\nu$ denotes the order, $\mathrm{K}_{\nu}(\cdot)$ and $\mathrm{\Gamma}(\cdot)$ refer to the modified Bessel and Gamma functions, respectively. The kernel hyperparameters --- $\sigma_f$, $l$, and $\alpha$ --- control the output scale, correlation length, and scale mixture of kernel components, respectively. We adopt a Bayesian approach, training these hyperparameters by marginalizing over the log-likelihood,
\begin{equation}
\ln \mathcal{L}(\sigma_f, l, \alpha, \ldots) = -\frac{1}{2} D^\top (\mathcal{K} + \mathcal{C})^{-1} D - \frac{1}{2} \ln |\mathcal{K} + \mathcal{C}| - \frac{N}{2} \ln (2\pi),
\end{equation}
where $D \equiv D(\{x_i\})$ is the observed data vector, $N$ is the size of the training data, $\mathcal{K} \equiv \{\kappa_{ij}\}$ is the GP kernel matrix evaluated on training points, and $\mathcal{C}$ is the noise covariance matrix. We assume uniform priors on the logarithms of hyperparameters (e.g., $\log_{10}\sigma_f \sim  \mathcal{U}[-5, 6]$, $\log_{10}l \sim  \mathcal{U}[-5, 6]$ and $\log_{10}\alpha \sim  \mathcal{U}[-5, 6]$), enabling flexible modelling over a wide range of behaviours.

Once trained, predictions at test points are made using the GP posterior mean and covariance:
\begin{align}
\overline{f^\star} &= \mathcal{K}^\star \left( \mathcal{K} + \mathcal{C} \right)^{-1} D \, \\
\mathrm{cov}[f^\star] &= \mathcal{K}^{\star\star} - \mathcal{K}^\star \left( \mathcal{K} + \mathcal{C} \right)^{-1} \left( \mathcal{K}^\star \right)^\top,
\end{align}
where $\mathcal{K}^\star$ is the cross-kernel between training and test points, and $\mathcal{K}^{\star\star}$ is the kernel evaluated at test points.

\item \textbf{Knot-based spline interpolation:} As an alternative, flexible technique, we implement free-form spline interpolation defined over a set of fixed nodes or ``knots''. Let the function of interest $f(x)$ be represented by a spline of order $k$ (with $k=3$, $4$, or $5$ tested in our analysis), defined over $n$ knot locations $\{x_1, x_2, \ldots, x_n\}$, with corresponding function values $\{f_1, f_2, \ldots, f_n\}$. The function is constructed as a piecewise polynomial of degree $k-1$, with continuity and smoothness conditions imposed up to the $(k-2)$-th derivative at each knot location. These knot locations are selected to balance flexibility, informed by features identified in analyses.

The knot values ${f_m}$ are treated as free parameters and inferred by fitting the spline to observational data $D \equiv D(\{x_i\})$, with uncertainties encoded in a noise covariance matrix $\mathcal{C}$. The fit is obtained by maximizing the Gaussian log-likelihood:
\begin{equation}
\ln \mathcal{L} = -\frac{1}{2} D^\mathrm{T} \, \mathcal{C}^{-1} \, D - \frac{1}{2} \ln \left| \mathcal{C} \right| - \frac{N}{2} \ln(2\pi),
\end{equation}
where $D = y_{\rm obs} - f(x_{\rm obs}; {x_m, f_m})$ is the residual between the observed data and the spline-interpolated function evaluated at the same locations.

We perform Bayesian parameter inference, utilizing Markov Chain Monte Carlo (MCMC) sampling, to explore the posterior distribution of the spline coefficients $\{f_m\}$. The reconstructed function and its derivatives—evaluated analytically from the spline representation—are thus obtained as posterior distributions, allowing reliable propagation of observational uncertainties. For related applications of free-form or knot-based spline reconstruction in cosmological contexts, one can refer to \cite{Brumback01091998, Sealfon:2005em, Sahni:2006pa, Planck:2015sxf, Zhao:2017cud}. 
\end{itemize}
Together, these two reconstruction strategies---GPs and knot-based splines---offer a robust, cosmology-independent framework for recovering smooth functions and their derivatives from noisy, discrete observations.

\section{Observational Datasets \label{sec:data}}

Our analysis is based on two key datasets: the DESI DR2 BAO measurements and the DES-5YR SN data. These complementary probes provide geometric information across a wide redshift range, which we use to reconstruct cosmological distance measures and their derivatives.
\begin{itemize}
    \item The DES-SN 5YR \cite{DES:2024jxu} sample provides redshift-dependent measurements of the distance modulus $\mu(z)$, which is related to the luminosity distance $d_L(z)$ through the expression:
\begin{equation}
\mu(z) = m_B(z) - M_B=  5 \log_{10} \left[ \frac{d_L(z)}{\text{Mpc}} \right] + 25 \, .
\end{equation}
Here, $m_B$ is the SN-Ia apparent magnitude, and $M_B$ is the absolute magnitude of the SN-Ia. In our reconstruction framework, $M_B$ enters as a nuisance parameter in the distance modulus and directly affects the normalization of $d_L(z)$. This allows us to reconstruct $d_L(z)$ in a self-consistent way, properly accounting for the calibration uncertainty.
\item The DESI DR2 \cite{DESI:2025zgx} provides high-precision measurements of the BAO feature, which constrain the comoving distance $d_M(z)$ and the Hubble distance $d_H(z) = c/H(z)$, both normalized by the sound horizon at the drag epoch, $r_d$. Thus, the BAO observables are $\frac{d_M(z)}{r_d}$ and $\frac{d_H(z)}{r_d}$ respectively.
For reconstruction purposes, to recover $d_M(z)$ and $d_H(z)$ from the DESI BAO observables, either $r_d$ is fixed to a reference cosmological value (e.g., Planck 2018) or treated as a nuisance parameter to be marginalized over in the statistical inference.
\item  In addition to low-redshift probes, we incorporate a compressed constraint from the cosmic microwave background (CMB) through the angular scale of the sound horizon at recombination, $\theta_\star \approx r_\star / d_M(z_\star)$, where $r_\star$ is the comoving sound horizon at recombination and $d_M(z_\star)$ is the transverse comoving distance to the recombination redshift $z_\star$. Assuming standard $\Lambda$CDM physics prior to recombination, we use the relation $r_\star \approx r_d / 1.0187$ and adopt $z_\star \approx 1090$ following Ref. \cite{Planck:2018vyg}. We impose a Gaussian prior on $100\,\theta_\star$, with a mean value of $1.0411$ and a standard deviation of $0.0005$ (approximately twice the Planck 2018 reported uncertainty). This constraint serves as a high-redshift geometric anchor, enhancing sensitivity to potential deviations from Planck $\Lambda$CDM. Our choice is motivated by ACT results \cite{ACT:2025fju, ACT:2025tim}, indicating that the early universe is consistent with, and shows no evidence for new physics, beyond Planck $\Lambda$CDM. 
\end{itemize}

\section{Reconstructing Cosmic Distance-Derivatives \label{sec:recon}}

We now proceed to reconstruct the comoving distance $d_M(z)$ and its derivative \textit{w.r.t} redshift, $d_M'(z)$, in a model-independent manner. These quantities are fundamental to inferring various cosmological distance measures and the Hubble parameter at different redshifts, without assuming a specific background model. We employ both MTGP and knot-based spline interpolation for this exercise. Both approaches are designed to incorporate uncertainties and propagate them into derived quantities consistently.

From the reconstructed $d_M(z)$, we can derive the angular diameter distance $d_A(z) = d_M(z)/(1+z)$ and the luminosity distance $d_L(z) = (1+z) d_M(z)$, along with their corresponding derivatives \textit{w.r.t} redshift. These derivatives are crucial to extract the Hubble parameter at characteristic redshifts $z_i \, \text{ for } \, i = 1, \,2, \,3,  \ldots , 7$, identified through geometric distance identities given in Eq. \eqref{eq:z_char}, directly utiliizng Eq. \eqref{eq:Hz_char}, independent of specific cosmological models. For representation, we derive the distances in dimensionless fashion independent of $H_0$, and redefine the normalised distances and their derivatives as,
\begin{align}
&    D_{M}(z) = \frac{H_0}{c} d_M(z) \,  ; \quad \,  D'_{M}(z) = \frac{H_0}{c}d'_M(z)\, ,\nonumber \\
&    D_{A}(z) = \frac{H_0}{c} d_A(z) \,  ; \quad \,  D'_{A}(z) = \frac{H_0}{c}d'_A(z)\, , \\
&    D_{L}(z) = \frac{H_0}{c} d_L(z) \,  ; \quad \,  D'_{L}(z) = \frac{H_0}{c}d'_L(z)\, . \nonumber
\end{align}
Similarly, we can represent the Hubble parameter [Eq. \eqref{eq:Hz_char}] in a dimensionless form $E(z) = {H(z)}/{H_0}$, at the characteristic redshifts, as 
\begin{align} \label{eq:Ez_char}
    & E(z_1) = \frac{1}{D_A(z_1) \left[ (1+z_1)^{3} + 1\right]}  \, , \nonumber \\ % z4
    & E(z_2) = \frac{1}{D_A(z_2) \left[ (1+z_1)^{2} + 1\right]}  \, , \nonumber\\ % z3
    & E(z_3) = \frac{1}{D_A(z_3) \left[ 2 + z_3\right]}  \, , \nonumber\\ % z2
    & E(z_4) = \frac{1}{D_A(z_4) \left[ (1+z_4)^{2} \right]}  \, , \\ % z7
    & E(z_5) = \frac{1}{D_A(z_5) \left[ 1+z_1 \right]}  \, , \nonumber\\ % z5
    & E(z_6) = \frac{1}{D_A(z_6) \, z_6}  \, , \nonumber\\ % z8
    & E(z_7) = \frac{1}{D_A(z_7)}  \, . \nonumber % z1 and z6
\end{align}

Our analysis rests on two foundational assumptions: (i) we assume the validity of the cosmic distance-duality relation (CDDR), which connects the luminosity and angular diameter distances via $d_L(z) = (1+z)^2 d_A(z)$ and holds in any metric theory of gravity with photon number conservation; (ii) we assume no significant anomalies or outliers between the datasets, implying that the two observational probes are mutually consistent with no systematic biases. These assumptions are empirically testable, and we will revisit them in subsequent sections where we assess the internal consistency of the reconstructed quantities.
\begin{itemize}
    \item The MTGP framework enables the joint reconstruction of correlated cosmological functions---in this case, $D_M(z)$ and its derivative $D_M'(z)$---while capturing their covariances and uncertainties. We implement the reconstruction using the \texttt{tinygp} \cite{tinygp}, \texttt{jax} \cite{jax2018github} and \texttt{NumPyro} \cite{numpyro, bingham2019pyro} packages. We test the stability of our findings for different kernel choices: Mat\'{e}rn-$\nu$ class with $\nu=7/2$ (M72) and $9/2$ (M92), the squared exponential (RBF), and the rational quadratic kernel (RQ), respectively.
    
    \begin{figure}[t]
    \centering
    \includegraphics[width=0.495\linewidth]{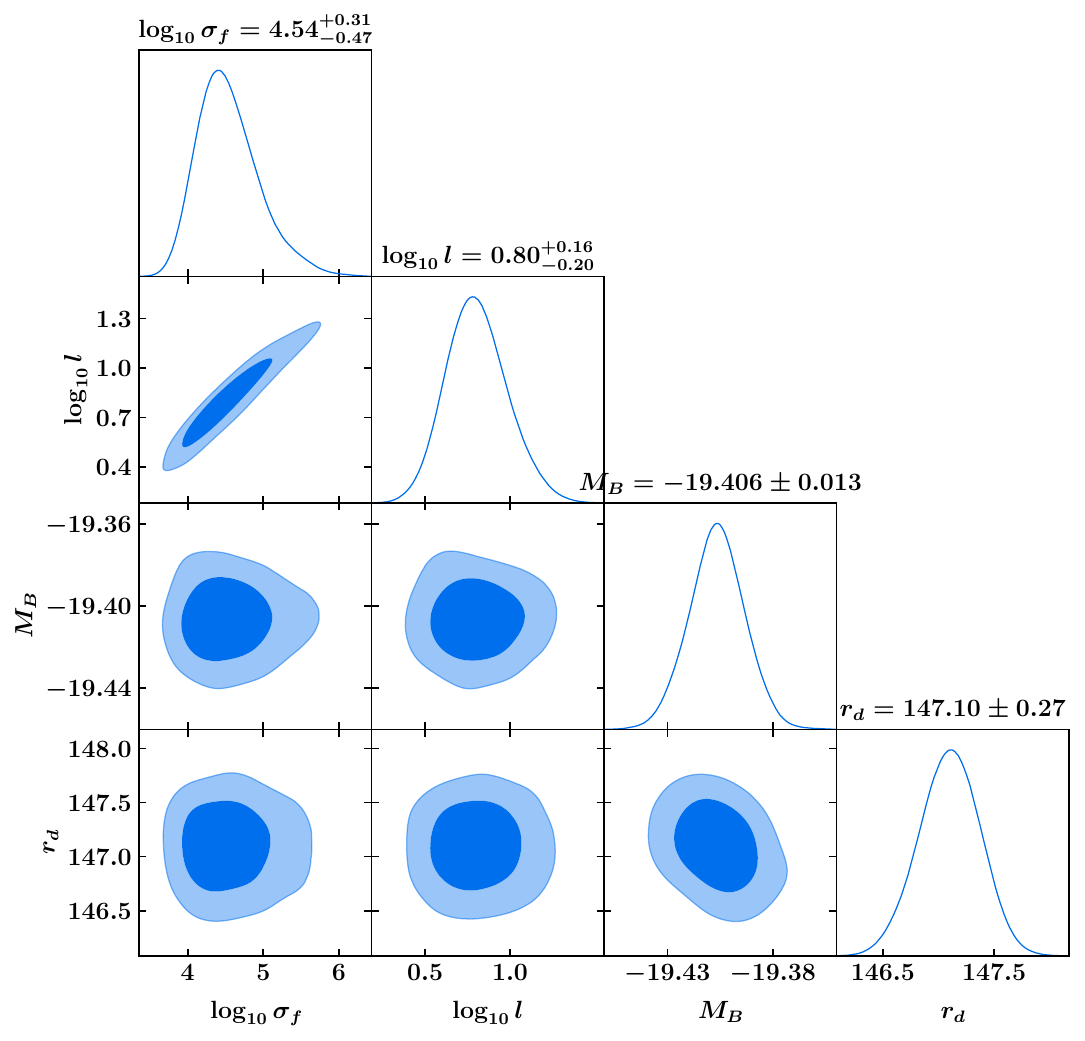}
    \includegraphics[width=0.495\linewidth]{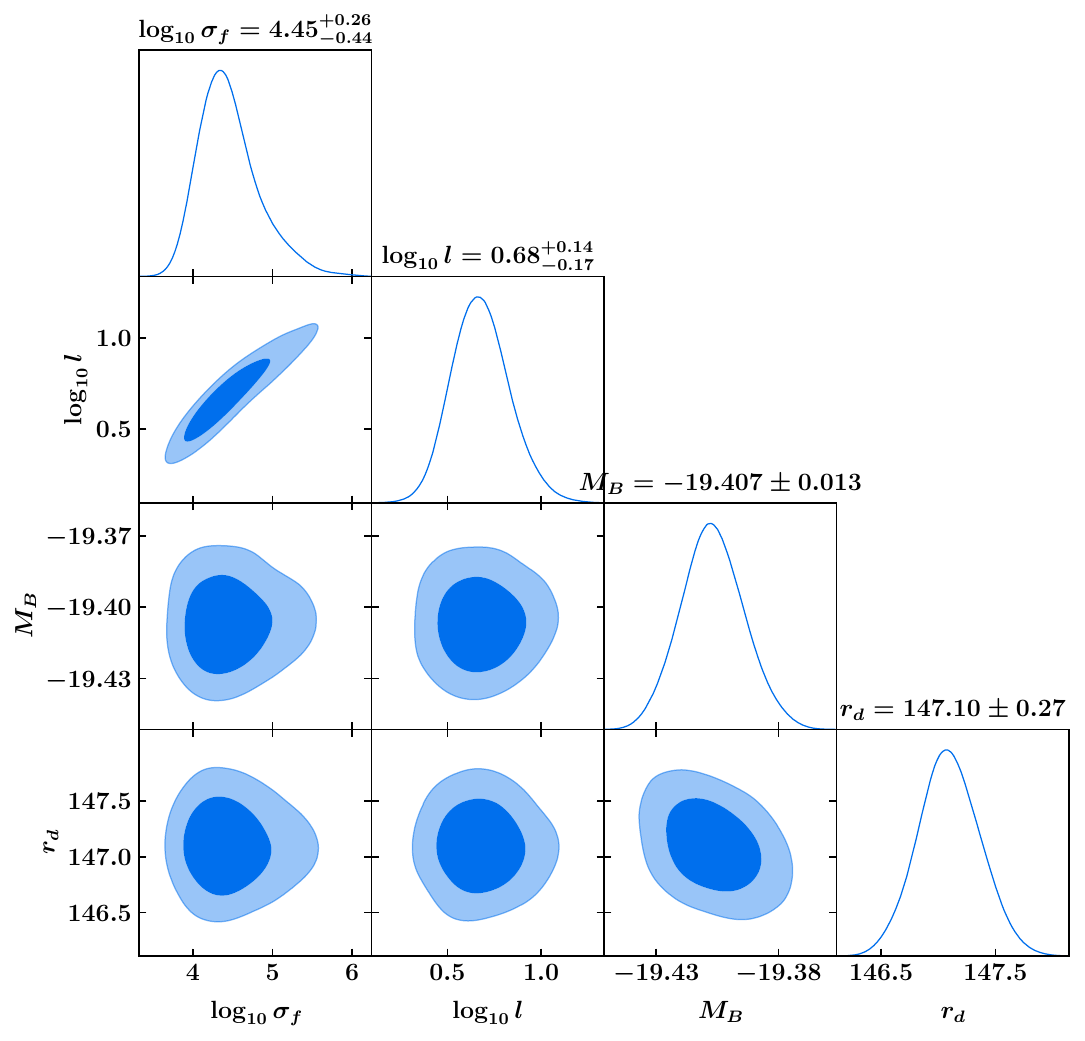} \\
    \includegraphics[width=0.495\linewidth]{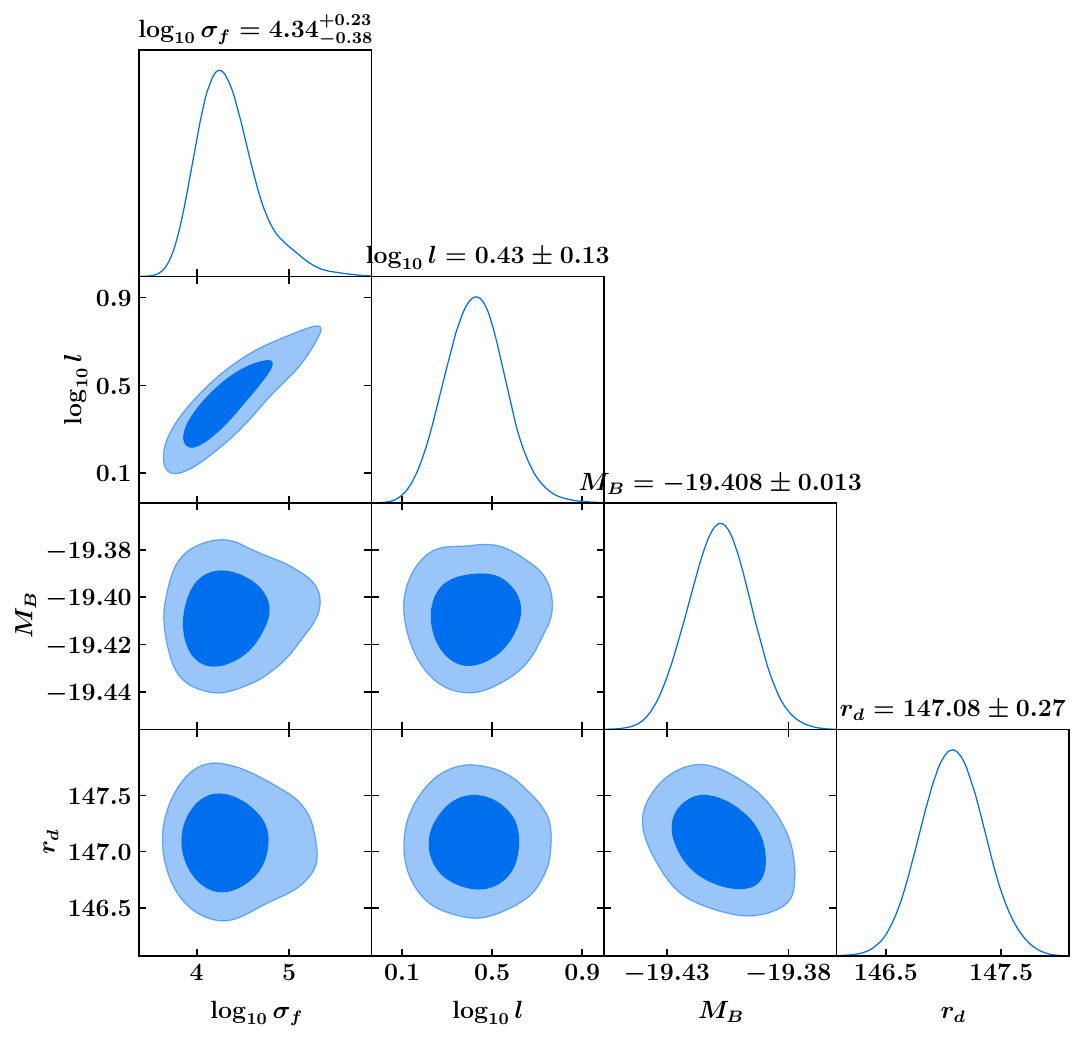}
    \includegraphics[width=0.495\linewidth]{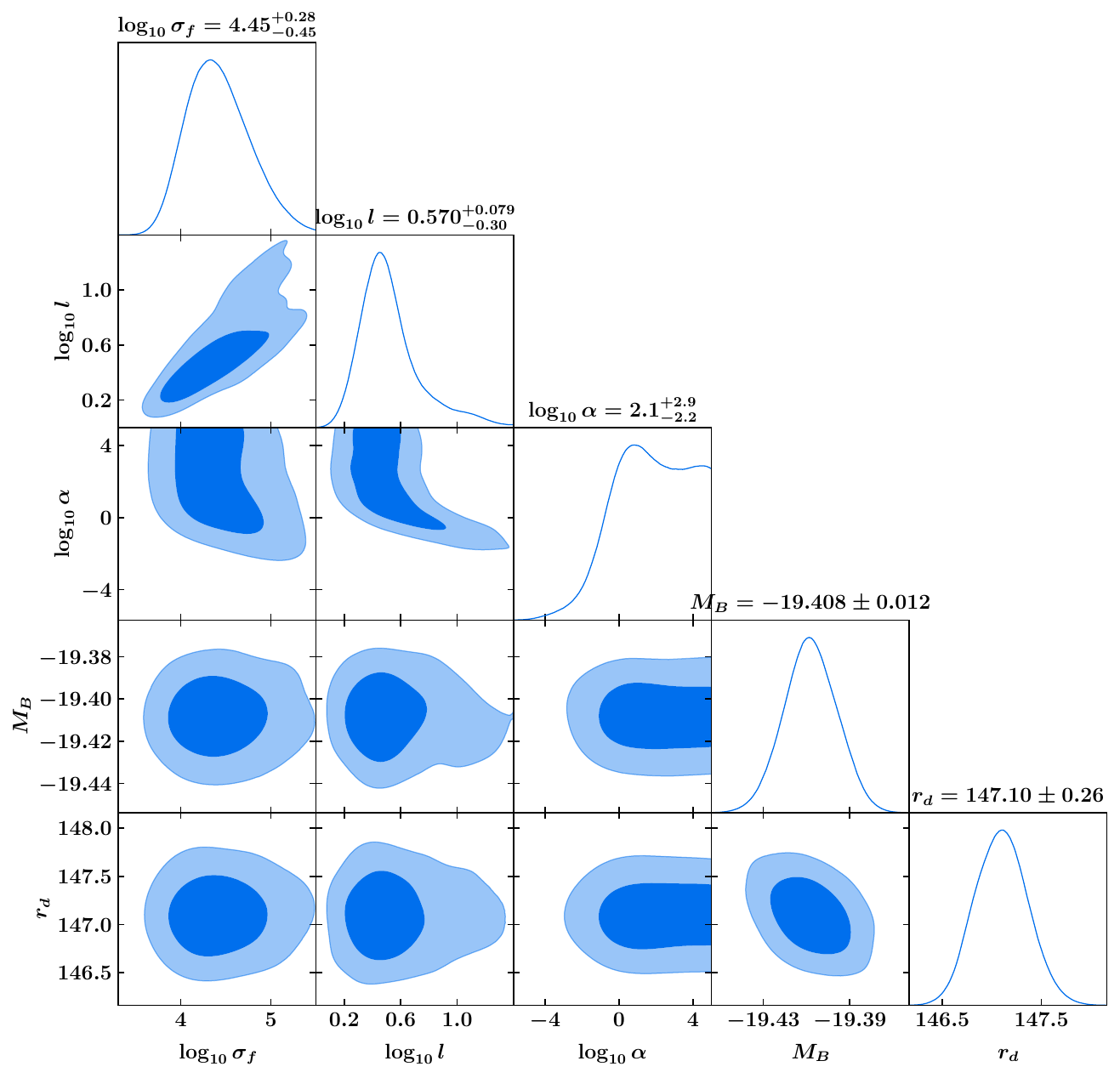}
    \caption{Triangle plot for MTGP hyperparameters, with marginalised constraints on $M_B$ and $r_d$ (in Mpc), assuming (i) M72 kernel [top left], (ii) M92 kernel [top right], (iii) RBF kernel [bottom left], (iv) RQD kernel [bottom right].}
    \label{fig:gp_mcmc}
\end{figure}

\begin{figure}[t]
    \centering
    \includegraphics[width=0.9\linewidth]{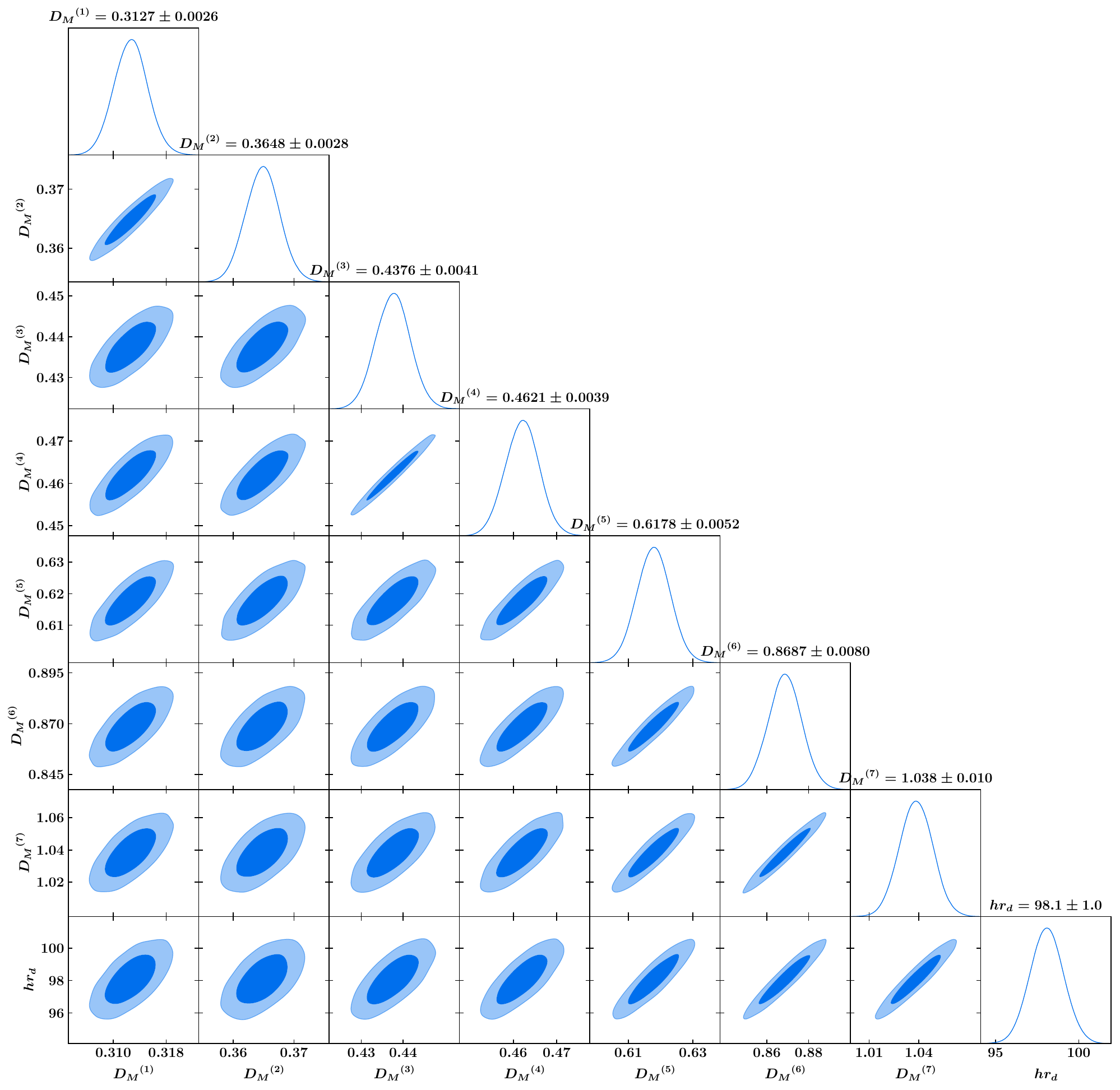} 
    \caption{Triangle plot for the $D_M^{(i)}$, $i=1, \ldots,7$, at the characteristic redshift knots, assuming Zero Mean function along with marginalised constraints on $hr_d$ [in units of km/s].}
    \label{fig:rec_mcmc}
\end{figure}

\begin{figure}[t]
    \centering
    \includegraphics[width=0.9\linewidth]{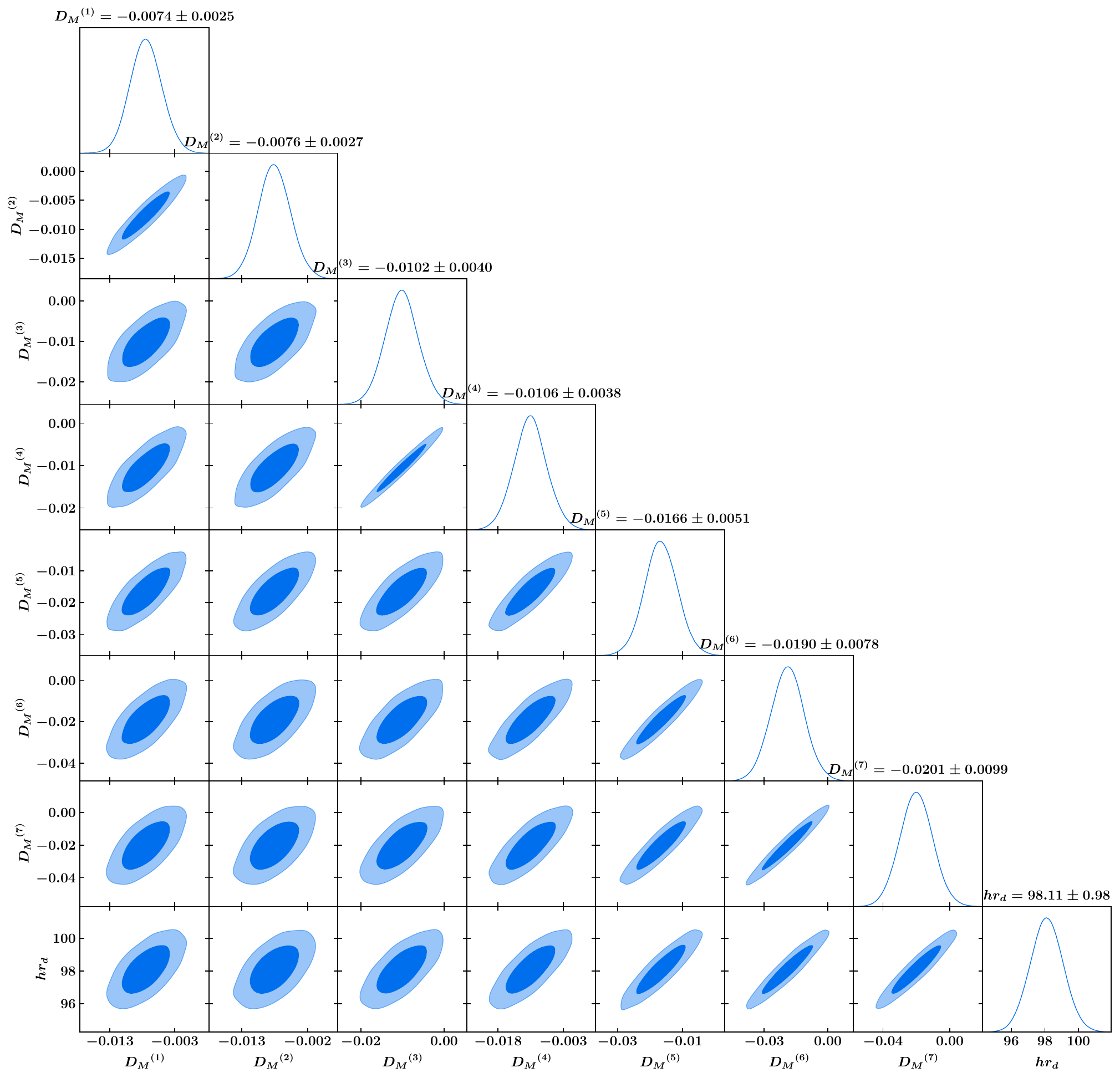} 
    \caption{Triangle plot for the $D_M^{(i)}$, $i=1, \ldots,7$, at the characteristic redshift knots, assuming Planck 2018 $\Lambda$CDM best-fit mean function along with marginalised constraints on $hr_d$ [in units of km/s].}
    \label{fig:residual_mcmc}
\end{figure}

\item We adopt uniform, uninformative priors on all kernel hyperparameters, as discussed in Sec. \ref{sec:method}. The MTGP is trained on the combined DESI DR2 and DES 5YR datasets, where we simultaneously vary the sound horizon at drag epoch \cite{Eisenstein:1997ik}, $r_d$, with a Gaussian prior in the range $r_d \in \mathcal{N}(147.09, 0.27)$~Mpc \cite{Planck:2018vyg}, and assume a uniform prior over the SN-Ia absolute magnitude $M_B \in \mathcal{U}[-21, -18]$. This ensures consistent marginalization over uncertainties in the BAO standard ruler and SN-Ia calibration. Using the marginalized posteriors of the kernel hyperparameters, $M_B$ and $r_d$, we reconstruct $D_M(z)$ and $D_M'(z)$, and propagate these into predictions for $D_A(z)$, $D_L(z)$, and their derivatives.
    \item From the reconstructed $D_M(z)$, $D_A(z)$, $D_L(z)$, and their derivatives \textit{w.r.t.} redshift, we identify the characteristic redshifts, i.e., $z_i \, \text{ for } \, i = 1, \,2, \,3,  \ldots , 7$. At each of these redshifts, we compute $D_A(z_i)$ and hence $E(z_i)$, utilizing Eq. \eqref{eq:Ez_char} respectively. 
    \item  In the second part of our analysis, we use the characteristic redshifts $\{z_i\}$ identified from the MTGP analysis as nodes (or ``knots'') for a knot-based spline interpolation of $D_M(z)$ across these specific redshifts. This free-form approach provides an intuitive, localized view of the cosmic expansion history by directly fitting the values of $D_M$ at the knot positions. We explore three different spline orders, $k = 3, 4, 5$, to assess the stability of the reconstruction under changes in smoothness. The interpolation is performed either directly on the $D_M$ values (with a zero mean) or on residuals defined \textit{w.r.t} the Planck 2018 baseline $\Lambda$CDM predictions. 
    \item  To incorporate high-redshift CMB information into our residual-based approach, we include an additional knot at $z = z_\star = 1090$, corresponding to the redshift of recombination. At $z_\star$, we constrain the residual of the comoving distance with respect to $\Lambda$CDM by matching the angular scale of the first acoustic peak in the CMB, $\theta_\star \approx r_\star / d_M(z_\star)$. Assuming standard pre-recombination physics, we relate the comoving sound horizon at recombination to that at the drag epoch via $r_\star \approx r_d / 1.0187$. We then impose a Gaussian prior on $100\,\theta_\star$ with a mean of $1.0411$ and standard deviation $0.0005$, consistent with Planck 2018 \cite{Planck:2018vyg} but with slightly inflated uncertainty to remain conservative. This high-redshift anchor helps tie down the behavior of $d_M(z)$ at early times and improves the overall stability and accuracy of the reconstruction, especially in scenarios allowing for deviations from $\Lambda$CDM at lower redshifts.
    \item  For each case, we adopt flat, uninformative priors on the $D_M(z_i)$ values ($D_M(z_i) \in [0, 10]$ for zero mean or $D_M(z_i) \in [-10, 10]$ for residuals defined \textit{w.r.t} Planck 2018 baseline). In addition, we marginalize over the SN-Ia absolute magnitude $M_B$, and adopt a flat prior on $hr_d \in \mathcal{U}[80,120] \, \text{km s}^{-1}$, where $h = H_0/(100 \, \text{km} \, \text{s}^{-1} \, \text{Mpc}^{-1})$ is the dimensionless Hubble constant. We perform parameter inference with \texttt{Cobaya} \cite{Torrado:2020dgo} to derive the posterior distributions of the $D_M$ values at the knots, which we interpolate across the range of redshift jointly probed by both the data sets to get our final smoothed predictions. 
    \item  The reconstructed $D_M(z)$ function and its derivative $D_M'(z)$ are then used to compute $D_A(z)$, $D_A'(z)$, $D_L(z)$, and $D_L'(z)$, respectively. Henceforth, we again identify the characteristic redshifts, yielding an independent set of geometric derivative crossing points, employing knot-based spline interpolation, complementary to the MTGP reconstruction framework. 
\end{itemize}

\begin{figure}[t]
    \centering
    \includegraphics[width=0.325\linewidth]{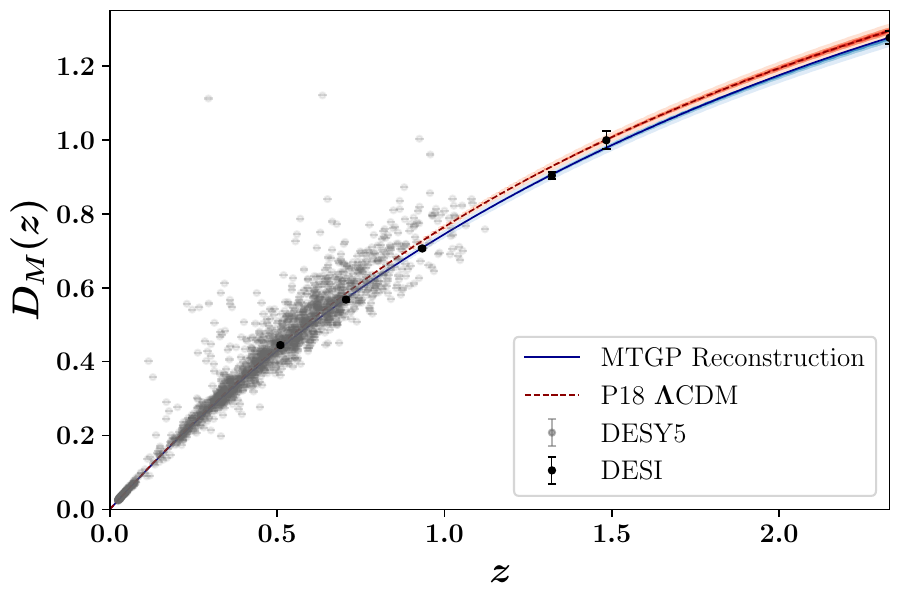}
    \includegraphics[width=0.325\linewidth]{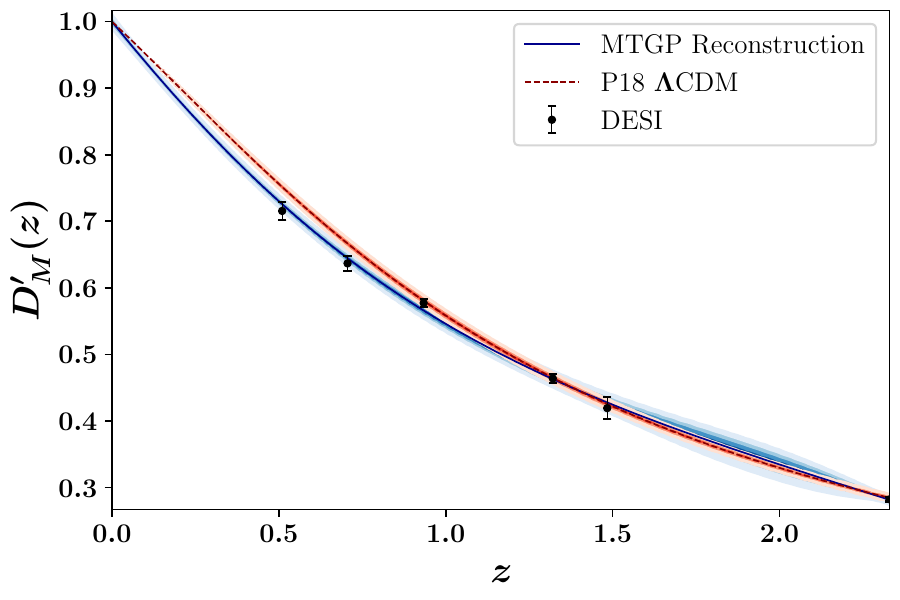}
    \includegraphics[width=0.325\linewidth]{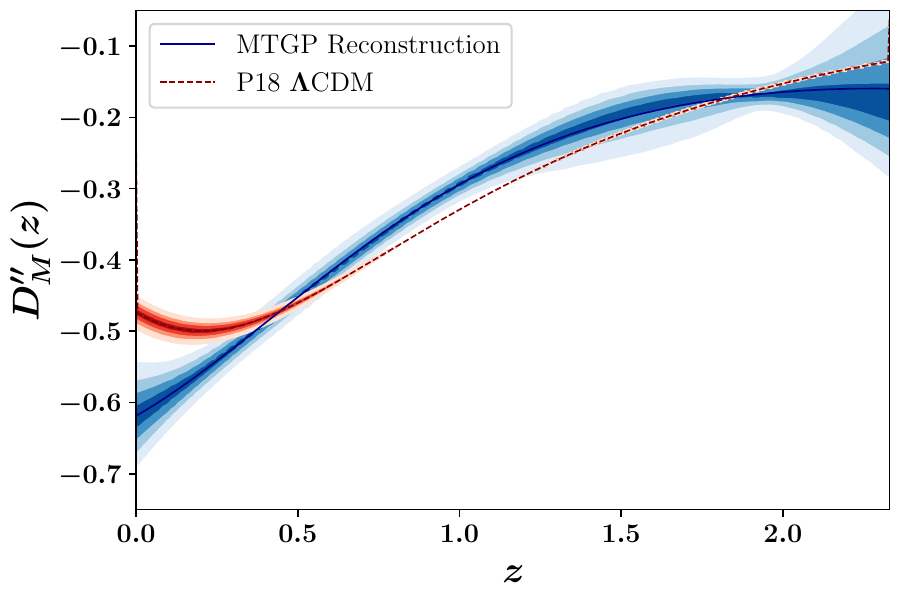} \\
    \includegraphics[width=0.325\linewidth]{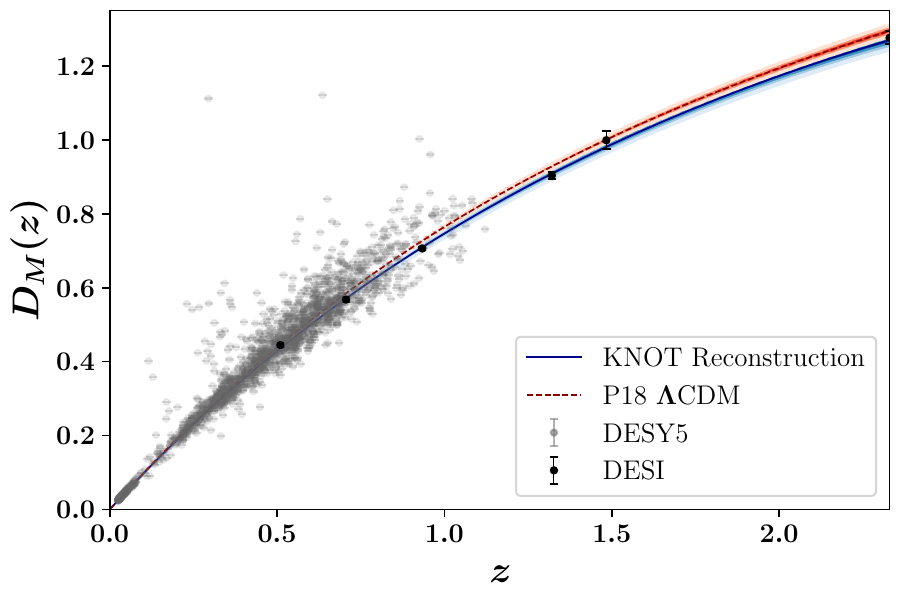}
    \includegraphics[width=0.325\linewidth]{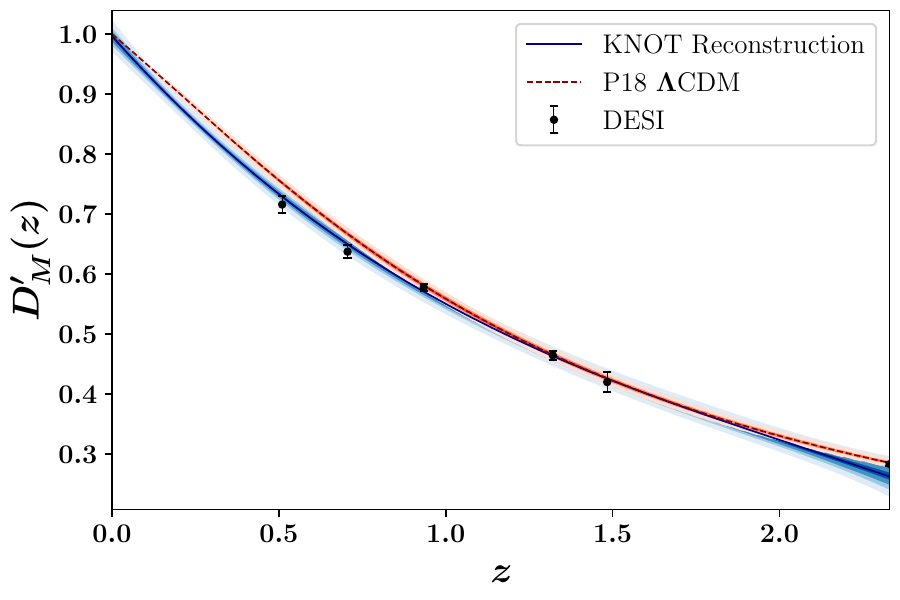}
    \includegraphics[width=0.325\linewidth]{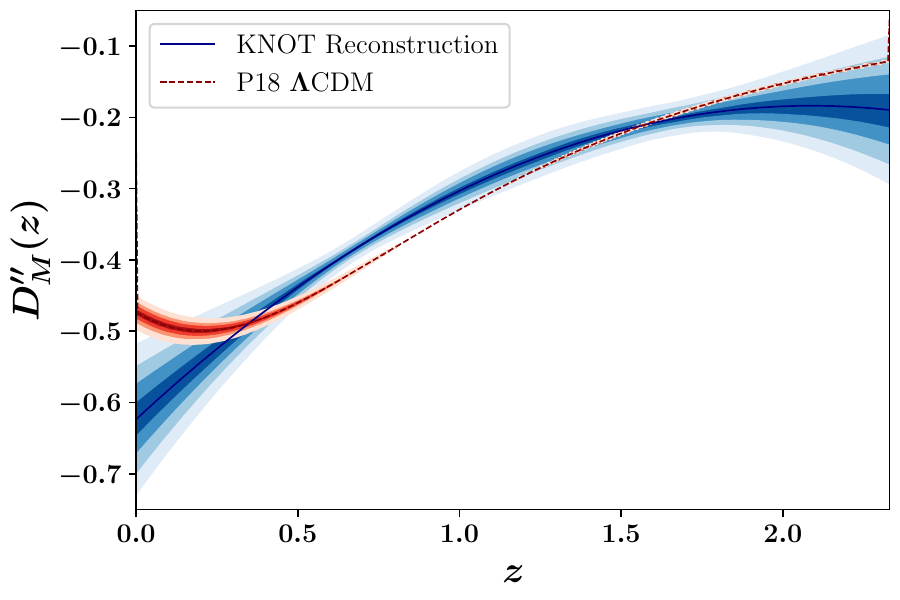}
    \caption{Reconstruction of $D_M(z)$, $D_M'(z)$ and $D_M''(z)$, employing (i) MTGP reconstruction with RQD kernel [in upper panel], and (ii) Free-form Knot-based spline reconstruction with $k=4$ order [in lower panel] respectively.}
    \label{fig:DM_rec}
\end{figure}

 These characteristic redshifts $\{z_i\}$ are obtained by solving the condition $D_{\rm \{X\}}(z_i) = D'_{\rm \{Y\}}(z_i)$, where two pairs of the reconstructed function ($\mathrm{\{X\}} \text{ and } \mathrm{\{Y\}} \equiv M, A \text{ or } L$) and the distance derivatives intersect, as defined in Eq. \eqref{eq:z_char}. To estimate the uncertainty in each crossing point, we propagate the errors on $D_{\rm \{X\}}(z)$ and $D'_{\rm \{Y\}}(z)$ and determine the bounds of the crossing using the conditions:
\begin{equation}
D_{\rm \{X\}}(z) + \sigma_{D_{\rm \{X\}}}(z) = D_{\rm \{Y\}}'(z) - \sigma_{D'_{\rm \{Y\}}}(z) \, , \quad
D_{\rm \{X\}}(z) - \sigma_{D_{\rm \{X\}}}(z) = D_{\rm \{Y\}}'(z) + \sigma_{D'_{\rm \{Y\}}}(z) \, .
\end{equation}    
Since $D_{\rm \{X\}}(z)$ is monotonic increasing function (note: $D_L$ is always increasing \textit{w.r.t} $z$, whereas $D_A$ is an increasing function up to $z=z_7$ --- where it has a maxima --- after which it decreases for $z>z_7$) with $0$ value at present epoch, while $D'_{\rm \{Y\}}(z)$ is a decreasing function whose value is unity as $z=0$, these conditions yield a pair of redshift values, $z_{i, \text{left}}$ and $z_{i, \text{right}}$, which define the $1\sigma$ uncertainty range around each characteristic $z_i$. We define the central value as the point of intersection and compute the uncertainty as $\sigma_{z_i} = (z_{i, \text{right}} - z_{i, \text{left}})/2$.

Using these bounds, we compute the corresponding distance measures $D_{\rm \{X\}}(z)$ at $z_{i, \text{left}}$ and $z_{i, \text{right}}$, from which the uncertainty on $D_{\rm \{ X\}}(z_i)$ is inferred as:
\begin{equation}
    \sigma_{D_{\rm \{X\}}(z_i)} = \frac{1}{2} \left| D_{\rm \{X\}}(z_{i, \text{right}}) - D_{\rm \{X\}}(z_{i, \text{left}}) \right| \, .
\end{equation}   
Finally, we can compute the normalized Hubble parameter $E(z_i)$ at each $z_i$, using the analytic relations in Eq.~\eqref{eq:Ez_char}. The uncertainty on $E(z_i)$, $\sigma_{E}(z_i)$ is then computed by standard error propagation. Thus, our approach offers a robust, model-independent way to extract both the values and uncertainties in the characteristic redshifts, inferred distances, and expansion rate estimates, i.e., $\{z_i, \, D_{\rm \{X\}}(z_i), \, E(z_i)\}$, thereby testing consistency of the expansion history across two reconstruction frameworks. 
\medskip

Having outlined the full reconstruction and inference methodology in detail, we now turn to presenting the key results derived from our analysis, summarized below:
\begin{itemize}[left=0pt]
    \item  Figure~\ref{fig:gp_mcmc} presents the triangle plots for the MTGP hyperparameters and the marginalized posteriors of the nuisance parameters $M_B$ and $r_d$ (in Mpc), derived using four different kernel choices: M72, M92, RBF, and RQD, with \texttt{GetDist} \cite{Lewis:2019xzd}. These plots illustrate the correlations between the kernel hyperparameters ($\log_{10} \sigma_f$, $\log_{10} l$, $\log_{10} \alpha$) and the astrophysical or cosmological parameters ($M_B$, $r_d$ in Mpc). The stability of $M_B$ and $r_d$ posteriors across kernels demonstrates the robustness of the GP reconstructions to different smoothness assumptions. 
    \item  Figure~\ref{fig:rec_mcmc} and Figure~\ref{fig:residual_mcmc} present the joint posterior distributions for the dimensionless comoving distances $D_M^{(i)}$ at the seven characteristic redshift knots ${z_i}$, obtained using a knot-based spline interpolation assuming a zero mean baseline, and residuals defined relative to the Planck 2018 $\Lambda$CDM best-fit mean function. Also shown is the marginalized constraint on $hr_d$ in km s$^{-1}$ for both cases, for comparison. 
    \item  To evaluate the performance of both the reconstruction methods, we compute the reduced chi-squared statistic, $\chi^2_{\rm min}/\mathrm{d.o.f}$, where $\mathrm{d.o.f} = 1735 + 13 - n$, accounts for the total degrees of freedom $\equiv$ total number of BAO and SN data points - number of parameters ($n$) in the reconstruction algorithm. Both reconstruction frameworks yield statistically consistent fits to the data, with only minor variations in the $\chi^2_{\rm min}/\mathrm{d.o.f}$ across different kernel choices and spline orders. Specifically, we obtain $\chi^2/\mathrm{d.o.f} = 0.88$ for the MTGP reconstruction with RQD kernel, and $\chi^2_{\rm min}/\mathrm{d.o.f} = 0.94$ for the knot-based interpolator with spline order $k = 4$. Both reconstruction frameworks deliver comparably good fits to the data, with only slight variations in $\chi^2_{\rm min}/\mathrm{d.o.f}$ across different kernels and spline orders. These results indicate excellent agreement with the data across methods, suggesting that the reconstructions are statistically equivalent in performance.

\begin{figure}[t]
    \centering
    \includegraphics[width=0.495\linewidth]{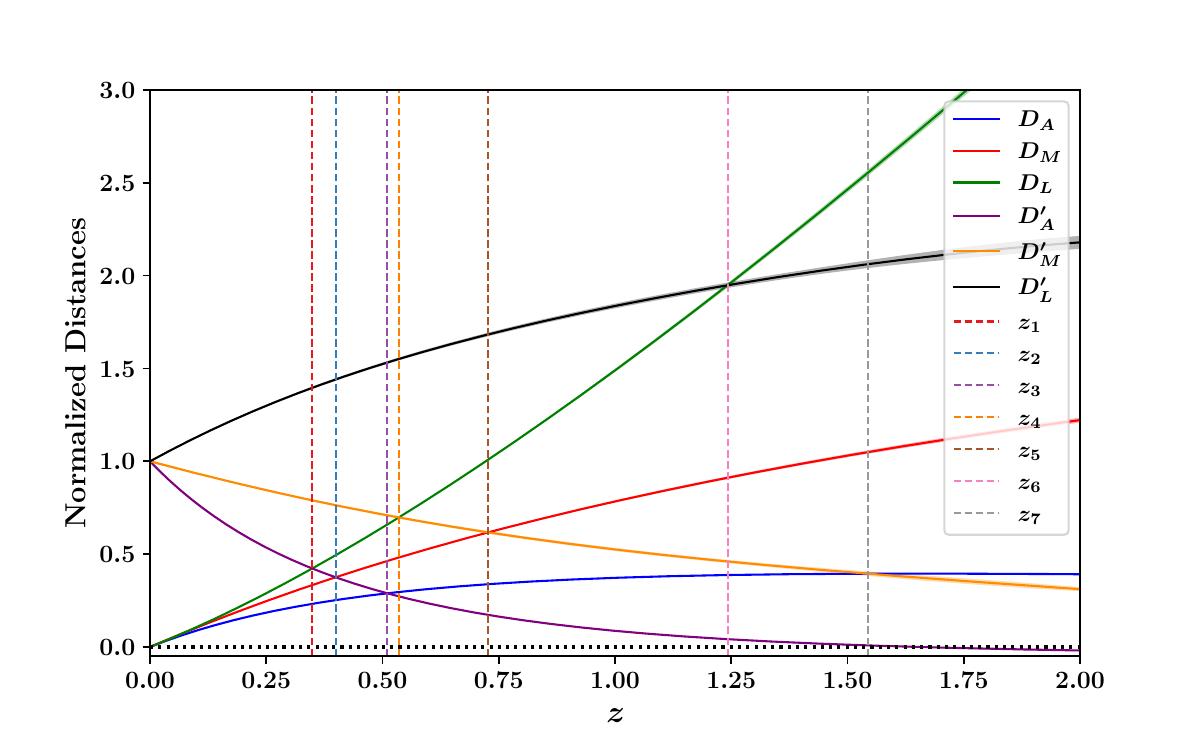}
    \includegraphics[width=0.495\linewidth]{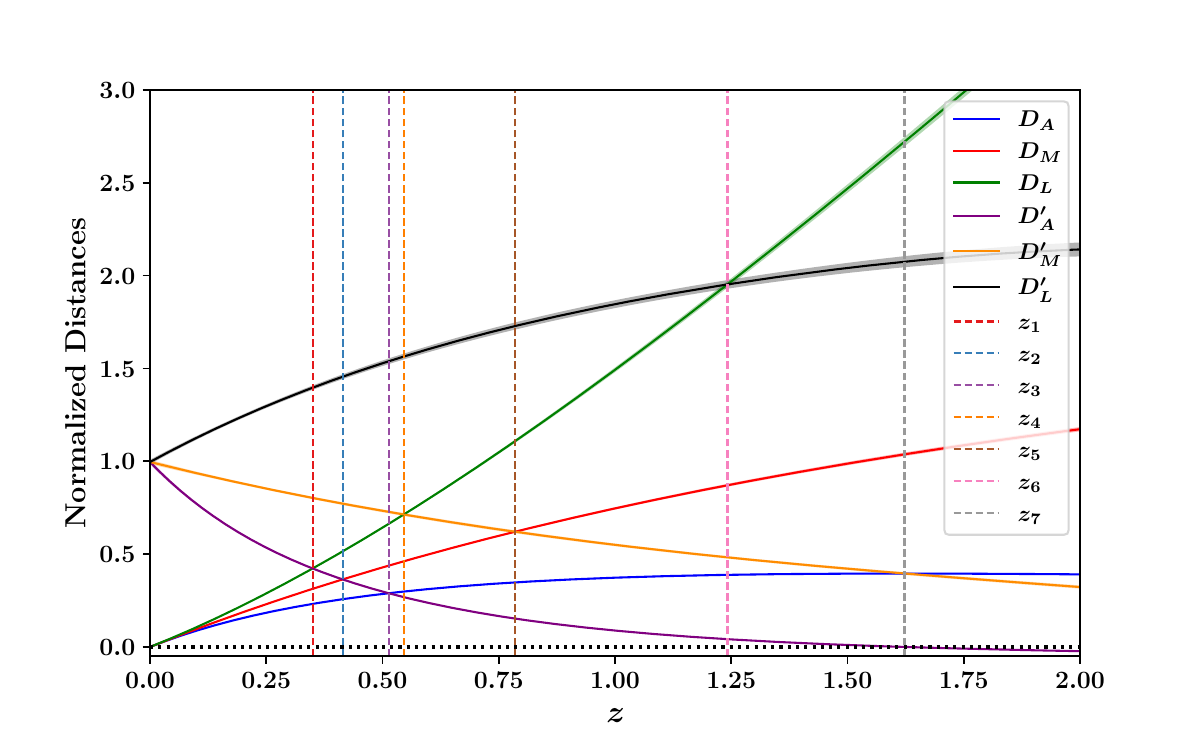}
    \caption{Evolution of the normalized cosmic distances: $d_A(z)$, $d_M(z)$ and $d_L(z)$, and their derivatives ($d_A'(z)$, $d_M'(z)$ and $d_L'(z)$), employing (i) MTGP reconstruction with RQD kernel [in left panel], and (ii) Free-form Knot-based spline reconstruction with $k=4$ order [in right panel], $z_i$ ($i = 1 \cdots 7$) are the characteristic redshifts. The plots include $1\sigma$ uncertainties, although they are not visibly distinguishable here.}
    \label{fig:z_rec}
\end{figure}

\begin{figure}[t]
    \centering
    \includegraphics[width=0.49\linewidth]{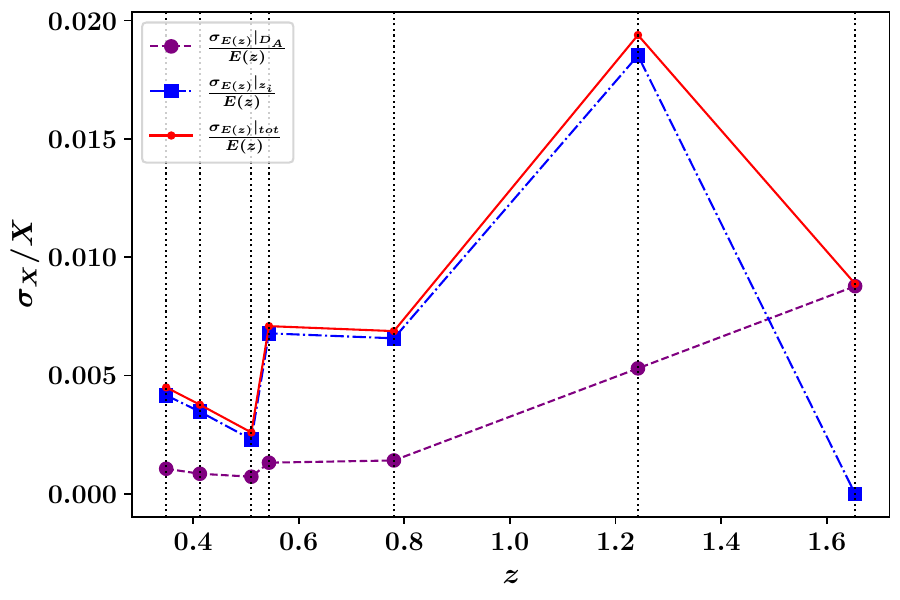}
    \includegraphics[width=0.49\linewidth]{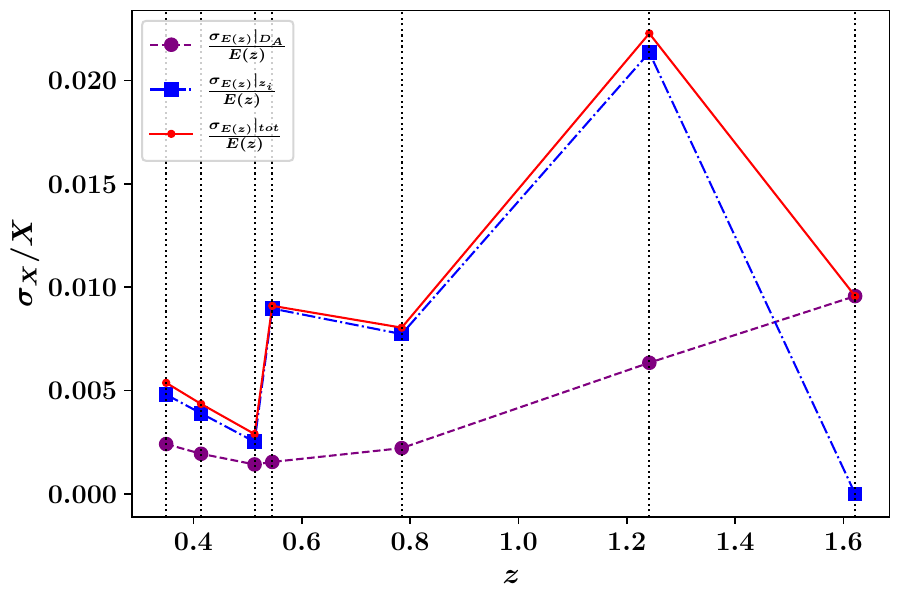}
    \caption{Relative uncertainties in the reconstructed normalized expansion rate $E(z)$ at each of the 7 characteristic redshifts $z_i$. The quantity $\sigma_{E\vert_{D_A}}(z_i)/E(z_i)$ (purple circles, dashed lines) denotes the fractional contribution to the uncertainty in $E(z_i)$ due to the error in the angular diameter distance $D_A(z_i)$. The term $\sigma_{E\vert_z}(z_i)/E(z_i)$ (blue squares, dash-dotted lines) represents the contribution from uncertainties in the redshift $z_i$. The total fractional uncertainty $\sigma_{E_\vert{\rm tot}}(z_i)/E(z_i)$ (red triangles, solid lines) includes both sources. The left and right panels correspond to reconstructions using MTGP framework with RQD kernel, and spine-based knots with $k=4$ for zero mean function. Vertical dashed lines mark the positions of the identified characteristic redshifts.}
    \label{fig:err_contrib}
\end{figure}

\begin{table*}
\centering
\caption{Reconstructed values of $D_A(z)$ and $E(z)$ at the characteristic redshifts $z_i$, obtained from MTGP method for different kernel choices. The Planck 2018 $\Lambda$CDM predictions are also shown, along with the tension in units of $\sigma$.}
\renewcommand{\arraystretch}{1.25}
\setlength{\tabcolsep}{6pt}
\begin{tabular}{ccccccc}
\toprule
\multicolumn{7}{c}{\textbf{Mat\'ern $\nu=7/2$ (M72) kernel}} \\
\hline
$z_i$ & $D_A(z_i)$ & $D_A^{\text{P18}}(z_i)$ & Tension ($\sigma$) 
      & $E(z_i)$ & $E^{\text{P18}}(z_i)$ & Tension ($\sigma$) \\
\midrule
$z_1$=0.349 $\pm$ 0.003  & 0.232 $\pm$ 0.001 & 0.236 $\pm$ 0.001 & 3.3  & 1.247 $\pm$ 0.006 & 1.208 $\pm$ 0.005 & 5.35 \\
$z_2$=0.412 $\pm$ 0.004  & 0.257 $\pm$ 0.001 & 0.262 $\pm$ 0.002 & 3.31 & 1.299 $\pm$ 0.005 & 1.254 $\pm$ 0.006 & 5.62 \\
$z_3$=0.510 $\pm$ 0.007  & 0.289 $\pm$ 0.001 & 0.295 $\pm$ 0.002 & 3.14 & 1.380 $\pm$ 0.004 & 1.331 $\pm$ 0.009 & 5.22 \\
$z_4$=0.543 $\pm$ 0.006  & 0.298 $\pm$ 0.001 & 0.305 $\pm$ 0.002 & 3.83 & 1.409 $\pm$ 0.011 & 1.358 $\pm$ 0.009 & 3.59 \\
$z_5$=0.782 $\pm$ 0.014  & 0.347 $\pm$ 0.001 & 0.356 $\pm$ 0.003 & 3.63 & 1.619 $\pm$ 0.013 & 1.571 $\pm$ 0.017 & 2.22  \\
$z_6$=1.244 $\pm$ 0.024  & 0.388 $\pm$ 0.002 & 0.397 $\pm$ 0.002 & 3.01  & 2.073 $\pm$ 0.042 & 2.061 $\pm$ 0.034 & 0.21  \\
$z_7$=1.628 $\pm$ 0.094  & 0.395 $\pm$ 0.004 & 0.403 $\pm$ 0.002 & 1.83 & 2.531 $\pm$ 0.024 & 2.532 $\pm$ 0.124 & 0.01 \\

\hline
\multicolumn{7}{c}{\textbf{Mat\'ern $\nu=9/2$ (M92) kernel}} \\
\hline
$z_i$ & $D_A(z_i)$ & $D_A^{\text{P18}}(z_i)$ & Tension ($\sigma$) 
      & $E(z_i)$ & $E^{\text{P18}}(z_i)$ & Tension ($\sigma$) \\
\midrule
$z_1$=0.349 $\pm$ 0.003  & 0.232 $\pm$ 0.001 & 0.236 $\pm$ 0.001 & 3.41 & 1.247 $\pm$ 0.005 & 1.208 $\pm$ 0.005 & 5.45 \\
$z_2$=0.413 $\pm$ 0.004  & 0.257 $\pm$ 0.001 & 0.262 $\pm$ 0.002 & 3.44 & 1.298 $\pm$ 0.005 & 1.254 $\pm$ 0.006 & 5.73\\
$z_3$=0.510 $\pm$ 0.006  & 0.289 $\pm$ 0.001 & 0.296 $\pm$ 0.002 & 3.34 & 1.380 $\pm$ 0.003 & 1.331 $\pm$ 0.008 & 5.36 \\
$z_4$=0.544 $\pm$ 0.006  & 0.298 $\pm$ 0.001 & 0.305 $\pm$ 0.002 & 4.01 & 1.408 $\pm$ 0.010 & 1.358 $\pm$ 0.009 & 3.71 \\
$z_5$=0.781 $\pm$ 0.013  & 0.347 $\pm$ 0.001 & 0.356 $\pm$ 0.002 & 3.81 & 1.619 $\pm$ 0.012 & 1.571 $\pm$ 0.017 & 2.36 \\
$z_6$=1.243 $\pm$ 0.024  & 0.388 $\pm$ 0.002 & 0.397 $\pm$ 0.002 & 3.11 & 2.074 $\pm$ 0.041 & 2.061 $\pm$ 0.033 & 0.26 \\
$z_7$=1.635 $\pm$ 0.086  & 0.395 $\pm$ 0.004 & 0.403 $\pm$ 0.002 & 1.88 & 2.531 $\pm$ 0.024 & 2.541 $\pm$ 0.114 & 0.08 \\
\hline
\multicolumn{7}{c}{\textbf{Squared Exponential (RBF) kernel}} \\
\hline
$z_i$ & $D_A(z_i)$ & $D_A^{\text{P18}}(z_i)$ & Tension ($\sigma$) 
      & $E(z_i)$ & $E^{\text{P18}}(z_i)$ & Tension ($\sigma$) \\
\midrule
$z_1$=0.349 $\pm$ 0.003  & 0.232 $\pm$ 0.001 & 0.237 $\pm$ 0.001 & 3.36 & 1.246 $\pm$ 0.005 & 1.208 $\pm$ 0.005  & 5.27 \\
$z_2$=0.413 $\pm$ 0.004  & 0.257 $\pm$ 0.001 & 0.263 $\pm$ 0.002 & 3.41 & 1.297 $\pm$ 0.005 & 1.255 $\pm$ 0.006  & 5.55 \\
$z_3$=0.510 $\pm$ 0.006  & 0.289 $\pm$ 0.001 & 0.296 $\pm$ 0.002 & 3.38 & 1.378 $\pm$ 0.003 & 1.331 $\pm$ 0.008  & 5.29 \\
$z_4$=0.544 $\pm$ 0.005  & 0.298 $\pm$ 0.001 & 0.305 $\pm$ 0.002 & 3.98 & 1.407 $\pm$ 0.010 & 1.358 $\pm$ 0.009  & 3.72 \\
$z_5$=0.780 $\pm$ 0.012  & 0.347 $\pm$ 0.001 & 0.356 $\pm$ 0.002 & 3.89 & 1.620 $\pm$ 0.011 & 1.570 $\pm$ 0.016  & 2.63 \\
$z_6$=1.242 $\pm$ 0.024  & 0.387 $\pm$ 0.002 & 0.397 $\pm$ 0.002 & 3.24 & 2.077 $\pm$ 0.041 & 2.059 $\pm$ 0.033  & 0.34 \\
$z_7$=1.658 $\pm$ 0.086  & 0.395 $\pm$ 0.004 & 0.403 $\pm$ 0.002 & 1.85 & 2.531 $\pm$ 0.023 & 2.571 $\pm$ 0.115  & 0.34 \\

\hline
\multicolumn{7}{c}{\textbf{Rational Quadratic (RQD) kernel}} \\
\hline
$z_i$ & $D_A(z_i)$ & $D_A^{\text{P18}}(z_i)$ & Tension ($\sigma$) 
      & $E(z_i)$ & $E^{\text{P18}}(z_i)$ & Tension ($\sigma$) \\
\midrule
$z_1$=0.349 $\pm$ 0.003  & 0.232 $\pm$ 0.001 & 0.237 $\pm$ 0.001 & 3.41 & 1.246 $\pm$ 0.005 & 1.208 $\pm$ 0.005 & 5.35 \\
$z_2$=0.413 $\pm$ 0.004  & 0.257 $\pm$ 0.001 & 0.263 $\pm$ 0.002 & 3.47 & 1.297 $\pm$ 0.005 & 1.255 $\pm$ 0.006 & 5.63 \\
$z_3$=0.510 $\pm$ 0.006  & 0.289 $\pm$ 0.001 & 0.296 $\pm$ 0.002 & 3.46 & 1.378 $\pm$ 0.003 & 1.331 $\pm$ 0.008 & 5.36 \\
$z_4$=0.544 $\pm$ 0.005  & 0.298 $\pm$ 0.001 & 0.305 $\pm$ 0.002 & 4.08 & 1.407 $\pm$ 0.010 & 1.358 $\pm$ 0.008 & 3.78 \\
$z_5$=0.780 $\pm$ 0.012  & 0.347 $\pm$ 0.001 & 0.356 $\pm$ 0.002 & 3.98 & 1.620 $\pm$ 0.011 & 1.570 $\pm$ 0.016 & 2.64 \\
$z_6$=1.242 $\pm$ 0.023  & 0.388 $\pm$ 0.002 & 0.397 $\pm$ 0.002 & 3.27 & 2.077 $\pm$ 0.040 & 2.060 $\pm$ 0.032 & 0.33  \\
$z_7$=1.654 $\pm$ 0.086  & 0.395 $\pm$ 0.004 & 0.403 $\pm$ 0.002 & 1.89 & 2.531 $\pm$ 0.023 & 2.566 $\pm$ 0.114 & 0.3  \\
\bottomrule
\end{tabular} \label{tab:gp}
\end{table*}

    \begin{table*}
\centering
\caption{Reconstructed values of $D_A(z)$ and $E(z)$ at the characteristic redshifts $z_i$, obtained from the free-form knot-based method for spline order $k = 3$ for both zero-mean and Planck $\Lambda$CDM best-fit residual mean functions. The Planck 2018 $\Lambda$CDM predictions are also shown, along with the tension in units of $\sigma$.}
\renewcommand{\arraystretch}{1.25}
\setlength{\tabcolsep}{6pt}
\begin{tabular}{ccccccc}
\toprule
\multicolumn{7}{c}{$k=3$} \\
\hline
\multicolumn{7}{c}{\textbf{Zero Mean Function}} \\
\hline
$z_i$ & $D_A(z_i)$ & $D_A^{\text{P18}}(z_i)$ & Tension ($\sigma$) 
      & $E(z_i)$ & $E^{\text{P18}}(z_i)$ & Tension ($\sigma$) \\
\midrule
$z_1$=0.352 $\pm$ 0.003 & 0.232 $\pm$ 0.001 & 0.238 $\pm$ 0.001 & 3.94 & 1.242 $\pm$ 0.006 & 1.210 $\pm$ 0.005 & 4.01 \\
$z_2$=0.417 $\pm$ 0.004 & 0.258 $\pm$ 0.001 & 0.264 $\pm$ 0.002 & 3.9 & 1.290 $\pm$ 0.006 & 1.258 $\pm$ 0.006 & 3.83 \\
$z_3$=0.518 $\pm$ 0.005 & 0.291 $\pm$ 0.001 & 0.298 $\pm$ 0.002 & 3.78 & 1.367 $\pm$ 0.004 & 1.337 $\pm$ 0.008 & 3.21 \\
$z_4$=0.549 $\pm$ 0.006 & 0.299 $\pm$ 0.001 & 0.307 $\pm$ 0.002 & 3.68 & 1.392 $\pm$ 0.011 & 1.363 $\pm$ 0.009 & 2.0 \\
$z_5$=0.788 $\pm$ 0.012 & 0.349 $\pm$ 0.001 & 0.357 $\pm$ 0.002 & 3.23 & 1.602 $\pm$ 0.012 & 1.578 $\pm$ 0.016 & 1.2 \\
$z_6$=1.233 $\pm$ 0.026 & 0.389 $\pm$ 0.002 & 0.397 $\pm$ 0.002 & 2.47 & 2.085 $\pm$ 0.045 & 2.049 $\pm$ 0.035 & 0.63 \\
$z_7$=1.577 $\pm$ 0.052 & 0.395 $\pm$ 0.004 & 0.403 $\pm$ 0.002 & 1.87 & 2.532 $\pm$ 0.024 & 2.466 $\pm$ 0.070 & 0.88 \\
\hline
\multicolumn{7}{c}{\textbf{Planck 2018 $\Lambda$CDM best-fit Mean Function}} \\
\hline
$z_i$ & $D_A(z_i)$ & $D_A^{\text{P18}}(z_i)$ & Tension ($\sigma$) 
      & $E(z_i)$ & $E^{\text{P18}}(z_i)$ & Tension ($\sigma$) \\
\midrule
$z_1$=0.352 $\pm$ 0.003 & 0.232 $\pm$ 0.001 & 0.238 $\pm$ 0.001 & 4.08 & 1.242 $\pm$ 0.006 & 1.210 $\pm$ 0.005 & 4.11 \\
$z_2$=0.417 $\pm$ 0.004 & 0.258 $\pm$ 0.001 & 0.264 $\pm$ 0.001 & 4.03 & 1.290 $\pm$ 0.005 & 1.258 $\pm$ 0.006 & 3.91 \\
$z_3$=0.518 $\pm$ 0.005 & 0.291 $\pm$ 0.001 & 0.298 $\pm$ 0.002 & 3.89 & 1.367 $\pm$ 0.004 & 1.337 $\pm$ 0.008 & 3.24 \\
$z_4$=0.549 $\pm$ 0.006 & 0.299 $\pm$ 0.001 & 0.307 $\pm$ 0.002 & 3.79 & 1.392 $\pm$ 0.011 & 1.363 $\pm$ 0.009 & 2.03 \\
$z_5$=0.789 $\pm$ 0.012 & 0.349 $\pm$ 0.001 & 0.357 $\pm$ 0.002 & 3.32 & 1.602 $\pm$ 0.011 & 1.578 $\pm$ 0.016 & 1.21 \\
$z_6$=1.233 $\pm$ 0.025 & 0.389 $\pm$ 0.002 & 0.397 $\pm$ 0.002 & 2.5 & 2.085 $\pm$ 0.044 & 2.049 $\pm$ 0.034 & 0.63 \\
$z_7$=1.577 $\pm$ 0.051 & 0.395 $\pm$ 0.004 & 0.403 $\pm$ 0.002 & 1.88 & 2.531 $\pm$ 0.024 & 2.466 $\pm$ 0.069 & 0.88 \\
\bottomrule
\end{tabular} \label{tab:spline3}
\end{table*}

    \item For illustration purposes, we select two representative configurations that yield the most stable and slightly preferred results. Figure \ref{fig:DM_rec} presents the reconstructions of the comoving distance $D_M(z)$ (left column), its first derivative $D_M'(z)$ (middle column), and second derivative $D_M''(z)$ (right column) for these selected cases. The best-fit reconstruction curves are shown as dark blue solid lines, with the progressively lighter shaded regions denoting the 0.5$\sigma$, 1$\sigma$, 1.5$\sigma$, and 2$\sigma$ confidence levels (CLs), respectively, using the \texttt{fgivenx} \cite{fgivenx} package. For comparison, we also include the predictions from the baseline $\Lambda$CDM model, computed using the Planck 2018 TTTEEE+lensing chains. These are overlaid as red dashed lines, with uncertainties in red shaded regions, enabling a direct visual assessment of potential deviations from standard cosmology.

\begin{table*}
\centering
\caption{Reconstructed values of $D_A(z)$ and $E(z)$ at the characteristic redshifts $z_i$, obtained from the free-form knot-based method for spline order $k = 4$ for both zero-mean and Planck $\Lambda$CDM best-fit residual mean functions. The Planck 2018 $\Lambda$CDM predictions are also shown, along with the tension in units of $\sigma$.}
\renewcommand{\arraystretch}{1.25}
\setlength{\tabcolsep}{6pt}
\begin{tabular}{ccccccc}
\toprule
\multicolumn{7}{c}{$k=4$} \\
\hline
\multicolumn{7}{c}{\textbf{Zero Mean Function}} \\
\hline
$z_i$ & $D_A(z_i)$ & $D_A^{\text{P18}}(z_i)$ & Tension ($\sigma$) 
      & $E(z_i)$ & $E^{\text{P18}}(z_i)$ & Tension ($\sigma$) \\
\midrule
$z_1$=0.350 $\pm$ 0.003 &  0.232 $\pm$ 0.001  & 0.237 $\pm$ 0.001 & 3.23 & 1.246 $\pm$ 0.007  & 1.208 $\pm$ 0.005 & 4.52 \\
$z_2$=0.414 $\pm$ 0.004 &  0.257 $\pm$ 0.001  & 0.263 $\pm$ 0.002 & 3.34 & 1.296 $\pm$ 0.006  & 1.256 $\pm$ 0.006 & 4.77 \\
$z_3$=0.513 $\pm$ 0.006 &  0.289 $\pm$ 0.001  & 0.296 $\pm$ 0.002 & 3.34 & 1.375 $\pm$ 0.004  & 1.333 $\pm$ 0.009 & 4.34 \\
$z_4$=0.546 $\pm$ 0.007 &  0.299 $\pm$ 0.001  & 0.306 $\pm$ 0.002 & 3.39 & 1.402 $\pm$ 0.013  & 1.360 $\pm$ 0.009 & 2.67 \\
$z_5$=0.785 $\pm$ 0.014 &  0.348 $\pm$ 0.001  & 0.356 $\pm$ 0.003 & 3.34 & 1.612 $\pm$ 0.013  & 1.574 $\pm$ 0.017 & 1.74 \\
$z_6$=1.242 $\pm$ 0.027 &  0.388 $\pm$ 0.002  & 0.397 $\pm$ 0.002 & 2.64  & 2.073 $\pm$ 0.046  & 2.059 $\pm$ 0.036 & 0.24 \\
$z_7$=1.622 $\pm$ 0.056 &  0.396 $\pm$ 0.004  & 0.403 $\pm$ 0.002 & 1.73 & 2.528 $\pm$ 0.024  & 2.524 $\pm$ 0.077 & 0.05 \\
\hline
\multicolumn{7}{c}{\textbf{Planck 2018 $\Lambda$CDM best-fit Mean Function}} \\
\hline
$z_i$ & $D_A(z_i)$ & $D_A^{\text{P18}}(z_i)$ & Tension ($\sigma$) 
      & $E(z_i)$ & $E^{\text{P18}}(z_i)$ & Tension ($\sigma$) \\
\midrule
$z_1$=0.350 $\pm$ 0.003  &  0.232 $\pm$ 0.001 & 0.237 $\pm$ 0.001 & 3.32 & 1.246 $\pm$ 0.006 & 1.208 $\pm$ 0.005 & 4.6 \\
$z_2$=0.414 $\pm$ 0.004  &  0.257 $\pm$ 0.001 & 0.263 $\pm$ 0.002 & 3.43 & 1.295 $\pm$ 0.005 & 1.256 $\pm$ 0.006 & 4.82 \\
$z_3$=0.513 $\pm$ 0.006  &  0.289 $\pm$ 0.001 & 0.296 $\pm$ 0.002 & 3.42 & 1.375 $\pm$ 0.004 & 1.333 $\pm$ 0.008 & 4.46 \\
$z_4$=0.546 $\pm$ 0.007  &  0.299 $\pm$ 0.001 & 0.306 $\pm$ 0.002 & 3.51 & 1.402 $\pm$ 0.012 & 1.360 $\pm$ 0.009 & 2.72 \\
$z_5$=0.785 $\pm$ 0.013  &  0.348 $\pm$ 0.001 & 0.356 $\pm$ 0.002 & 3.43 & 1.611 $\pm$ 0.012 & 1.574 $\pm$ 0.017 & 1.76 \\
$z_6$=1.242 $\pm$ 0.026  &  0.389 $\pm$ 0.002 & 0.397 $\pm$ 0.002 & 2.68 & 2.072 $\pm$ 0.045 & 2.059 $\pm$ 0.035 & 0.23 \\
$z_7$=1.622 $\pm$ 0.056  &  0.396 $\pm$ 0.004 & 0.403 $\pm$ 0.002 & 1.74 & 2.528 $\pm$ 0.024 & 2.525 $\pm$ 0.076 & 0.03 \\
\bottomrule
\end{tabular} \label{tab:spline4}
\end{table*}

\begin{table*}
\centering
\caption{Reconstructed values of $D_A(z)$ and $E(z)$ at the characteristic redshifts $z_i$, obtained from the free-form knot-based method for spline order $k = 5$ for both zero-mean and Planck $\Lambda$CDM best-fit residual mean functions. The Planck 2018 $\Lambda$CDM predictions are also shown, along with the tension in units of $\sigma$.}
\renewcommand{\arraystretch}{1.25}
\setlength{\tabcolsep}{6pt}
\begin{tabular}{ccccccc}
\toprule
\multicolumn{7}{c}{$k=5$} \\
\hline
\multicolumn{7}{c}{\textbf{Zero Mean Function}} \\
\hline
$z_i$ & $D_A(z_i)$ & $D_A^{\text{P18}}(z_i)$ & Tension ($\sigma$) 
      & $E(z_i)$ & $E^{\text{P18}}(z_i)$ & Tension ($\sigma$) \\
\midrule
$z_1$=0.349 $\pm$ 0.004 & 0.232 $\pm$ 0.001 & 0.237 $\pm$ 0.002 & 2.67 & 1.248 $\pm$ 0.008 & 1.208 $\pm$ 0.005 &  4.23 \\
$z_2$=0.413 $\pm$ 0.005 & 0.257 $\pm$ 0.001 & 0.263 $\pm$ 0.002 & 2.75  & 1.298 $\pm$ 0.006 & 1.255 $\pm$ 0.007 &  4.54 \\
$z_3$=0.512 $\pm$ 0.008 & 0.289 $\pm$ 0.001 & 0.296 $\pm$ 0.002 & 2.92  & 1.376 $\pm$ 0.004 & 1.333 $\pm$ 0.009 &  4.35 \\
$z_4$=0.545 $\pm$ 0.008 & 0.298 $\pm$ 0.001 & 0.306 $\pm$ 0.002 & 3.18  & 1.403 $\pm$ 0.014 & 1.360 $\pm$ 0.010 &  2.56 \\
$z_5$=0.785 $\pm$ 0.014 & 0.348 $\pm$ 0.001 & 0.357 $\pm$ 0.003 & 3.33  & 1.611 $\pm$ 0.013 & 1.575 $\pm$ 0.017 & 1.69 \\
$z_6$=1.242 $\pm$ 0.027 & 0.389 $\pm$ 0.002 & 0.397 $\pm$ 0.002 &  2.64 & 2.072 $\pm$ 0.046 & 2.060 $\pm$ 0.036  & 0.2 \\
$z_7$=1.611 $\pm$ 0.079 & 0.396 $\pm$ 0.004 & 0.403 $\pm$ 0.002 & 1.71  & 2.528 $\pm$ 0.025 & 2.511 $\pm$ 0.105 & 0.16 \\
\hline
\multicolumn{7}{c}{\textbf{Planck 2018 $\Lambda$CDM best-fit Mean Function}} \\
\hline
$z_i$ & $D_A(z_i)$ & $D_A^{\text{P18}}(z_i)$ & Tension ($\sigma$) 
      & $E(z_i)$ & $E^{\text{P18}}(z_i)$ & Tension ($\sigma$) \\
\midrule
$z_1$=0.349 $\pm$ 0.004 & 0.232 $\pm$ 0.001 & 0.237 $\pm$ 0.002 & 2.71 & 1.247 $\pm$ 0.008 & 1.208 $\pm$ 0.005  & 4.22 \\
$z_2$=0.414 $\pm$ 0.005 & 0.257 $\pm$ 0.001 & 0.263 $\pm$ 0.002 & 2.76 & 1.297 $\pm$ 0.006 & 1.255 $\pm$ 0.007  & 4.48 \\
$z_3$=0.513 $\pm$ 0.007 & 0.289 $\pm$ 0.001 & 0.296 $\pm$ 0.002 & 2.95 & 1.376 $\pm$ 0.004 & 1.333 $\pm$ 0.009  & 4.31 \\
$z_4$=0.545 $\pm$ 0.007 & 0.298 $\pm$ 0.001 & 0.306 $\pm$ 0.002 & 3.24 & 1.403 $\pm$ 0.014 & 1.360 $\pm$ 0.010  & 2.6 \\
$z_5$=0.785 $\pm$ 0.014 & 0.348 $\pm$ 0.001 & 0.357 $\pm$ 0.002 & 3.4  & 1.611 $\pm$ 0.013 & 1.575 $\pm$ 0.017  & 1.7  \\
$z_6$=1.242 $\pm$ 0.026 & 0.389 $\pm$ 0.002 & 0.397 $\pm$ 0.002 & 2.67 & 2.071 $\pm$ 0.045 & 2.060 $\pm$ 0.035  & 0.21 \\
$z_7$=1.613 $\pm$ 0.081 & 0.396 $\pm$ 0.004 & 0.403 $\pm$ 0.002 & 1.72 & 2.528 $\pm$ 0.024 & 2.513 $\pm$ 0.107  & 0.13 \\
\bottomrule
\end{tabular} \label{tab:spline5}
\end{table*}

    \item The top row of Fig. \ref{fig:DM_rec} corresponds to the MTGP approach using the RQD kernel, while the bottom row shows the knot-based spline reconstruction with order $k=4$. These two methods exemplify complementary reconstruction strategies: MTGP leverages a global covariance structure that enforces smoothness across redshift, whereas the spline method offers local adaptability with minimal model assumptions. The first derivative $D_M'(z)$ directly relates to the Hubble expansion rate via $D_M'(z) = 1/E(z)$, while the second derivative $D_M''(z)$ reflects the redshift evolution of the expansion rate, which can potentially indicate departures from standard cosmological dynamics. Our findings demonstrate excellent agreement between the two approaches across the redshift range probed by the combined datasets. 

    \item The plots for reconstructed $D_M$, $D_M'$ and $D_M''$ (Fig. \ref{fig:DM_rec}, shown in blue) exhibit statistically significant deviations from the Planck $\Lambda$CDM prediction (shown in red), particularly at low to intermediate redshifts. In $D_M(z)$, the reconstructed curves lie consistently below the $\Lambda$CDM prediction for $z<1.5$, with the red curve falling outside the 2$\sigma$ blue bands in both methods. This discrepancy propagates to the first derivative $D_M'(z)$, where again the reconstructed values are higher relative to $\Lambda$CDM in the same redshift range. Most notably, in the second derivative $D_M''(z)$, both reconstructions show clear and strong deviations from $\Lambda$CDM at low redshift ($z \lesssim 0.5$), where the Planck curve lies well outside the $2\sigma$ uncertainty range. These consistent and statistically significant departures across multiple derivatives highlight the potential limitations of the baseline $\Lambda$CDM model and underscore the power of non-parametric reconstruction techniques in revealing subtle signatures of new physics in the redshift range of BAO and SN-Ia observations.

    \item With the reconstructed $D_M(z)$ and its derivative $D_M'(z)$, we can further derive related observables $D_A(z)$ and $D_L(z)$, along with their redshift derivatives. These quantities play a crucial role in locating the \textit{characteristic redshifts}. In particular, at the characteristic redshifts $\{z_i\}$, the expansion rate $E(z_i)$ can be directly computed using the reconstructed $D_A(z_i)$, without requiring any derivative information, or relying on any specific cosmological model assumptions, enabling a purely geometric and model-independent determination of $E(z_i)$ at those points. These distance-derivative crossing points are identified in Figure \ref{fig:z_rec} with all seven characteristic redshifts $z_1, \ldots, z_7$ marked. The left panel illustrates the results for the MTGP case, and the right panel shows the results from the knot-based reconstruction.

\item  The total uncertainty in the reconstructed expansion rate $E(z_i)$ at each characteristic redshift arises from the propagation of errors in both the angular diameter distance $D_A(z_i)$ and the characteristic redshift $z_i$, based on their analytic dependence in the model-independent expressions given in Eq. \eqref{eq:Ez_char}. Each of these relations has the form $E(z_i) = \left[ D_A(z_i) \cdot f(z_i) \right]^{-1}$, where $f(z_i)$ is a redshift-dependent algebraic factor. The uncertainty in $E(z_i)$, denoted $\sigma_{E}(z_i)$, is computed via standard error propagation as:
\begin{equation}
    \sigma_{E}^2(z_i) = \left\lbrace \frac{\partial E(z_i)}{\partial D_A(z_i)} \right\rbrace^2 \sigma_{D_A}^2(z_i) + \left\lbrace \frac{\partial E(z_i)}{\partial z_i} \right\rbrace^2 \sigma_z^2(z_i) \, ,
\end{equation}
where $\sigma_{D_A}(z_i)$ and $\sigma_z(z_i)$ represent the uncertainties in the distance and redshift, respectively. Since the dependence on $z_i$ enters nonlinearly through $f(z_i)$, the redshift-induced uncertainty contribution $\sigma_{E\vert_z}(z_i)$ varies significantly across different $z_i$. To visualize this decomposition, Fig. \ref{fig:err_contrib} shows the individual contributions $\sigma_{E\vert_{D_A}}(z_i)$, $\sigma_{E\vert_z}(z_i)$, and the total uncertainty $\sigma_{E\vert_{\rm tot}}(z_i)$ for all 7 $z_i$'s, across both MTGP (left panel) and knot-based (right panel) reconstructions. The results indicate that the uncertainty due to $z_{i}$ generally dominates, except for $z_{7}$ where $f(z_{7}) =1$. 
    
    \item We compare the computed $E(\{z_i\})$ and $D_A(\{z_i\})$ quantities to predictions of flat $\Lambda$CDM cosmology directly obtained from the Planck 2018 TTTEEE+lowE+lensing \cite{Planck:2018vyg} likelihood chains. The result of our analysis is summarized in Tables \ref{tab:gp}, \ref{tab:spline3}, \ref{tab:spline4} and \ref{tab:spline5}, for the MTGP and spline-based reconstruction frameworks. Finally, we derive the Gaussian tension metric between the reconstructed values vs Planck 2018 predictions. To this end, we account for the uncertainties associated with the characteristic redshifts $z_i$, i.e., $\sigma_{z_i}$ and propagate them into the Planck-inferred values of $D_A^{\rm P18}(z_i)$ and $E^{\rm P18}(z_i)$. This allows us to identify the redshift ranges where the derived results show noticeable deviations from the Planck 2018 baseline $\Lambda$CDM model.

    \item The reconstructed values of $D_A(\{z_i\})$ and $E(\{z_i\})$ at various redshifts show consistent deviations from the Planck 2018 $\Lambda$CDM predictions across all interpolation methods and kernel choices. For the MTGP method, we observe significant tensions in both $D_A({z})$ and $E({z_i})$, with deviations exceeding the $3\sigma$ level in $D_A(\{z_i\})$ and 5$\sigma$ level for $E(\{z_i\})$ at redshifts $z_1 \sim 0.35$, $z_2 \sim 0.41$, $z_3 \sim 0.51$, and $z_4 \sim 0.55$. These results persist across different kernel choices, with slight variations in the magnitude of the tension. Similarly, the knot-based reconstructions, using both zero-mean and Planck-based mean functions, reveal strong discrepancies in $H(z)$ exceeding $4\sigma$ around $z_1 \sim 0.35$, $z_2 \sim 0.41$, and $z_3 \sim 0.51$. In contrast, at higher redshifts such as $z_6 \sim 1.24$ and $z_7 \sim 1.6$, the reconstructions are consistent with Planck 2018 baseline $\Lambda$CDM expectations, with tensions in $E(\{z_i\})$ falling below $2\sigma$ and $D_A(\{z_i\})$ remaining within $3\sigma$. Crucially, this newly identified and statistically significant low-redshift discrepancy is not a manifestation of the well-known traditional $H_0$ tension \cite{DiValentino:2024yew}, but instead suggests the presence of additional deviations from the concordance model in the late-time Universe \cite{Vagnozzi:2023nrq}.

\begin{table*}
\centering
\caption{Summary of $D_A(z)$ and $E(z)$ measurements after systematically excluding individual tracer samples from DESI-DR2 BAO data. The top, middle, and bottom panels show results with LRG1, LRG2, and both LRG1+LRG2 tracers removed, respectively. The persistent $>3\sigma$ deviations in $E(z)$ at low redshifts ($z \sim 0.35$–$0.52$), despite excluding different subsets, indicate that the tensions are not driven by any single BAO data component, but may suggest a more fundamental departure from Planck  $\Lambda$CDM.}
\renewcommand{\arraystretch}{1.25}
\setlength{\tabcolsep}{6pt}
\begin{tabular}{ccccccc}
\toprule
\multicolumn{7}{c}{\textbf{No DESI-DR2 BAO LRG1 Tracers}} \\
\hline
$z_i$ & $D_A(z_i)$ & $D_A^{\text{P18}}(z_i)$ & Tension ($\sigma$) 
      & $E(z_i)$ & $E^{\text{P18}}(z_i)$ & Tension ($\sigma$) \\
\midrule
$z_1$= 0.352 $\pm$ 0.003 & 0.232 $\pm$ 0.001 & 0.238 $\pm$ 0.001 & 3.73 &  1.242 $\pm$ 0.007  &  1.210 $\pm$ 0.005  & 3.88 \\
$z_2$= 0.417 $\pm$ 0.004 &  0.258 $\pm$ 0.001 &  0.264 $\pm$ 0.002  & 3.61  &  1.290 $\pm$ 0.006  &  1.258 $\pm$ 0.006  & 3.8  \\
$z_3$= 0.518 $\pm$ 0.007  & 0.291 $\pm$ 0.001 & 0.298 $\pm$ 0.002  &  3.3 &  1.367 $\pm$ 0.004 & 1.337 $\pm$ 0.009  & 3.1  \\
$z_4$= 0.549 $\pm$ 0.007  & 0.299 $\pm$ 0.001  & 0.307 $\pm$ 0.002  & 3.37  &  1.392 $\pm$ 0.013  & 1.363 $\pm$ 0.009  & 1.8  \\
$z_5$= 0.791 $\pm$ 0.015  &  0.350 $\pm$ 0.001 &  0.357 $\pm$ 0.003  & 2.91  &  1.597 $\pm$ 0.013 &  1.580 $\pm$ 0.018  & 0.79 \\
$z_6$= 1.243 $\pm$ 0.028  & 0.391 $\pm$ 0.003  & 0.397 $\pm$ 0.002   &  1.96 & 2.059 $\pm$ 0.049  & 2.061 $\pm$ 0.038   & 0.02 \\
$z_7$= 1.610 $\pm$ 0.058  & 0.397 $\pm$ 0.004   &  0.403 $\pm$ 0.002  & 1.25  &  2.516 $\pm$ 0.025  &  2.509 $\pm$ 0.079  & 0.08 \\
\hline
\multicolumn{7}{c}{\textbf{No DESI-DR2 BAO LRG2 Tracers}} \\
\hline
$z_i$ & $D_A(z_i)$ & $D_A^{\text{P18}}(z_i)$ & Tension ($\sigma$) 
      & $E(z_i)$ & $E^{\text{P18}}(z_i)$ & Tension ($\sigma$) \\
\midrule
$z_1$=0.350 $\pm$ 0.003   & 0.232 $\pm$ 0.001  &  0.237 $\pm$ 0.001 & 3.29  & 1.245 $\pm$ 0.007  & 1.209 $\pm$ 0.005   & 4.44  \\
$z_2$=0.414 $\pm$ 0.004   & 0.257 $\pm$ 0.001  & 0.263 $\pm$ 0.002  &  3.3 &  1.295 $\pm$ 0.006 & 1.256 $\pm$ 0.006  & 4.64  \\
$z_3$=0.513 $\pm$ 0.007   & 0.290 $\pm$ 0.001  & 0.296 $\pm$ 0.002  & 3.17  & 1.374 $\pm$ 0.004  &  1.333 $\pm$ 0.009 & 4.24  \\
$z_4$=0.546 $\pm$ 0.007 & 0.299 $\pm$ 0.001  & 0.306 $\pm$ 0.002 & 3.33 &  1.401 $\pm$ 0.013  & 1.360 $\pm$ 0.009  &  2.52 \\
$z_5$=0.785 $\pm$ 0.014   & 0.348 $\pm$ 0.001  & 0.356 $\pm$ 0.003  & 3.26  & 1.611 $\pm$ 0.013   & 1.574 $\pm$ 0.018  & 1.69  \\
$z_6$=1.240 $\pm$ 0.027   & 0.388 $\pm$ 0.003  & 0.397 $\pm$ 0.002  & 2.63 &  2.077 $\pm$ 0.048  &  2.057 $\pm$ 0.037 & 0.34  \\
$z_7$=1.616 $\pm$ 0.058   &  0.395 $\pm$ 0.004 & 0.403 $\pm$ 0.002  & 1.79 &  2.530 $\pm$ 0.025  & 2.516 $\pm$ 0.078  &  0.17 \\
\hline
\multicolumn{7}{c}{\textbf{No DESI-DR2 BAO LRG1+LRG2 Tracers}} \\
\hline
$z_i$ & $D_A(z_i)$ & $D_A^{\text{P18}}(z_i)$ & Tension ($\sigma$) 
      & $E(z_i)$ & $E^{\text{P18}}(z_i)$ & Tension ($\sigma$) \\
\midrule
$z_1$=0.351 $\pm$ 0.003   & 0.232 $\pm$ 0.001 & 0.238 $\pm$ 0.001 &  3.61  & 1.243 $\pm$ 0.007 & 1.210 $\pm$ 0.005 & 3.97  \\
$z_2$=0.417 $\pm$ 0.005   & 0.258 $\pm$ 0.001  & 0.264 $\pm$ 0.002 &  3.46  & 1.291 $\pm$ 0.006 & 1.257 $\pm$ 0.006 & 3.89 \\
$z_3$=0.517 $\pm$ 0.007   & 0.290 $\pm$ 0.001 & 0.298 $\pm$ 0.002 & 3.09  & 1.368 $\pm$ 0.004 & 1.336 $\pm$ 0.009 & 3.21 \\
$z_4$=0.549 $\pm$ 0.008   & 0.299 $\pm$ 0.001 & 0.307 $\pm$ 0.002 &  3.24  & 1.394 $\pm$ 0.014  & 1.362 $\pm$ 0.010  & 1.85 \\
$z_5$=0.789 $\pm$ 0.016   & 0.349 $\pm$ 0.001 & 0.357 $\pm$ 0.003 & 2.83  & 1.601 $\pm$ 0.015 & 1.579 $\pm$ 0.019 & 0.91 \\
$z_6$=1.242 $\pm$ 0.032   & 0.390 $\pm$ 0.003 & 0.397 $\pm$ 0.002 & 2.04  & 2.066 $\pm$ 0.055 & 2.059 $\pm$ 0.041 & 0.11  \\
$z_7$=1.604 $\pm$ 0.067   & 0.397 $\pm$ 0.005 & 0.403 $\pm$ 0.002 & 1.29  & 2.521 $\pm$ 0.029 & 2.502 $\pm$ 0.090 & 0.2  \\
\hline
\end{tabular} \label{tab:desi_lrg}
\end{table*}

    \item  As part of our follow-up analysis, we conducted systematic tests to investigate whether the observed low-redshift tensions could be attributed to specific data subsets or known systematics. First, within the DESI BAO dataset, we removed individual tracer samples---specifically the LRG1 and LRG2 measurements---to assess their influence on the reconstruction \cite{Sapone:2024ltl, Colgain:2025nzf, Colgain:2024mtg, Wang:2024pui, Wang:2024rjd, Colgain:2024xqj, Carloni:2024zpl}. We found that the tension in the normalized expansion rate $E(z)$ at low redshifts ($z \sim 0.35$–$0.55$) persisted with deviations remaining above the $3\sigma$ CL, shown in Table. \ref{tab:desi_lrg}, irrespective of which subset was excluded (similar to the previous findings by \cite{Mukherjee:2024ryz, Mukherjee:2025fkf}). This robustness suggests that the observed discrepancies are not driven by any single component of the BAO dataset, and instead point toward a more fundamental deviation from $\Lambda$CDM predictions. 

\begin{table*}
\centering
\caption{Impact of applying a $\Delta \mu = -0.0482$ magnitude shift to low-redshift ($z \leq 0.1$) DES-SN5YR supernovae, following recent suggestions to account for uncorrected calibration systematics. Significant $>3\sigma$ tensions in $E(z)$ at $z \sim 0.35 - 0.51$ persist, indicating the discrepancy is not fully explained by low-redshift SN systematics.}
\renewcommand{\arraystretch}{1.25}
\setlength{\tabcolsep}{6pt}
\begin{tabular}{ccccccc}
\toprule
\multicolumn{7}{c}{\textbf{DES5 $\Delta \mu$=-0.0482 shift}} \\
\hline
$z_i$ & $D_A(z_i)$ & $D_A^{\text{P18}}(z_i)$ & Tension ($\sigma$) 
      & $E(z_i)$ & $E^{\text{P18}}(z_i)$ & Tension ($\sigma$) \\
\midrule
$z_1$=0.350 $\pm$ 0.004   & 0.233 $\pm$ 0.001  &  0.237 $\pm$ 0.002 &  2.26  & 1.241 $\pm$ 0.008   & 1.208 $\pm$ 0.005  &  3.53 \\
$z_2$=0.414 $\pm$ 0.006   & 0.258 $\pm$ 0.001  & 0.263 $\pm$ 0.002  &  2.14  & 1.291 $\pm$ 0.007   &  1.256 $\pm$ 0.007 & 3.55  \\
$z_3$=0.513 $\pm$ 0.009   & 0.291 $\pm$ 0.001  & 0.296 $\pm$ 0.003  &  2.03 & 1.370 $\pm$ 0.005   & 1.333 $\pm$ 0.010  &  3.22 \\
$z_4$=0.546 $\pm$ 0.010   & 0.300 $\pm$ 0.001  & 0.306 $\pm$ 0.003  & 2.2   &  1.397 $\pm$ 0.017  & 1.360 $\pm$ 0.011  &  1.79 \\
$z_5$=0.785 $\pm$ 0.020   & 0.349 $\pm$ 0.001  & 0.356 $\pm$ 0.003  &  2.12  & 1.606 $\pm$ 0.019   & 1.574 $\pm$ 0.022  &  1.09 \\
$z_6$=1.242 $\pm$ 0.037   & 0.390 $\pm$ 0.004  & 0.397 $\pm$ 0.002  &  1.62  & 2.064 $\pm$ 0.049   &  2.059 $\pm$ 0.046 & 0.07  \\
$z_7$=1.624 $\pm$ 0.061   & 0.397 $\pm$ 0.005  & 0.403 $\pm$ 0.002  &  1.05  & 2.518 $\pm$ 0.034   & 2.527 $\pm$ 0.082  &  0.1 \\
\hline
\end{tabular} \label{tab:des5_shift}
\end{table*}

    \item  In the DES-SN5YR supernova analysis, motivated by recent works \cite{Efstathiou:2024xcq, Berti:2025phi, Colgain:2024ksa}, we applied redshift offset corrections to the low-redshift ($z \leq 0.1$) SN Ia magnitudes to account for unknown calibration systematics. Despite this correction, the $E(z)$ parameter continued to show significant tension at $z \sim 0.35$ and $z \sim 0.41$, again exceeding the $3\sigma$ threshold, as demonstrated in Table \ref{tab:des5_shift}. These results indicate that even after mitigating potential low-redshift systematics in the DES-SN5YR supernova data, the deviations from Planck 2018 $\Lambda$CDM remain pronounced. This strengthens the case that the tension is unlikely to be a mere artefact of data imperfections but signals new physics in the late-time Universe. 
    
    \item  Despite the comprehensive tests undertaken in this study, we acknowledge that there may still be unknown systematics or unmodeled effects—\emph{unknown unknowns}—that could subtly bias the observables. These could arise from astrophysical contaminants, subtle selection effects, or cross-correlation errors that are not yet identified or fully understood. Given their elusive nature, a full propagation of such unknown systematics is beyond the scope of this study. However, we have made every effort to mitigate and account for all \emph{known unknowns} through independent reconstructions (Appendix. A), subset analyses and cross-validations (Appendix. B), and consistency checks (Sec. \ref{sec:anomalies}-\ref{sec:cddr}). The agreement across methods and datasets, and the persistence of anomalies under various stress tests, suggest that our main conclusions are not artifacts of known data imperfections. We believe this thorough treatment reinforces the robustness of our findings and supports their interpretation as potential signals of new physics in the late-time expansion history. 
\end{itemize}

\section{Evidence for New Physics? \label{sec:analysis}}

This section presents model-independent reconstruction of key cosmological functions derived from the joint analysis undertaken in Sec. \ref{sec:recon}, utilizing the reconstructed functions $D_M(z)$, $D_M'(z)$ and $D_M''(z)$ respectively.
\begin{itemize}[left=0pt]
\item From these quantities, we derive the normalized (dimensionless) Hubble parameter: $E(z) = 1/D_M'(z)$; the normalized expansion rate scaled with redshift: $E(z)/(1+z)$; and the  deceleration parameter: $q(z) = -1 + \frac{E'(z)}{E(z)}(1+z)$. We plot their evolution \textit{w.r.t.} redshift in Fig. \ref{fig:rec_set1}. The top panel corresponds to the results obtained with the MTGP framework with an RQD kernel, while the bottom panel is obtained with knot-based reconstruction, with spline order $k=4$. Best-fit curves from our reconstructions are shown in blue solid lines, while red dashed lines represent the Planck 2018 $\Lambda$CDM best-fit. Shaded bands indicate the 0.5$\sigma$, 1$\sigma$, 1.5$\sigma$, and 2$\sigma$ CL.
\item We observe significant deviations from the Planck $\Lambda$CDM model are observed at low redshifts ($z < 1$), with regions exceeding the 2$\sigma$ level. To better capture the late-time dynamics, we analyze $E(z)/(1+z)$, which is proportional to the expansion rate $\dot{a}$. The minimum of $E(z)/(1+z)$ corresponds to the redshift $z_t$ at which the Universe transitions from a decelerated to an accelerated phase of expansion, i.e., where $\ddot{a} = 0$. The transition redshift $z_t \lesssim 0.8$ suggests that the current datasets support an accelerated expansion phase below this $z_t$ within the 1$\sigma$ CL.
\item The slope of $E(z)/(1+z)$ near its minimum is gentler than in Planck $\Lambda$CDM. This means that (i) the increase in $\dot{a}$ with redshift is slower; (ii) the transition from a decelerating into the accelerating phase is more gradual; (iii) This behavior is described by the derivative, $\frac{d}{dz} \left[ \frac{E(z)}{1+z} \right] = \frac{E'(z)}{1+z} - \frac{E(z)}{(1+z)^2}$.
\item  The evolution of the deceleration parameter $q(z)$ also shows a zero-crossing at $z_t \lesssim 0.8$, consistent with the location of the $E(z)/(1+z)$ minimum. This reinforces the evidence for a late-time transition from a decelerated to accelerated expansion phase supported by the current datasets. Additionally, we find that the present-day value of the deceleration parameter $q_0 \equiv q(z=0)$ is less negative than the Planck $\Lambda$CDM prediction. This suggests a slowing down of cosmic acceleration in the late Universe, indicating deviations from a cosmological constant dark energy density and hinting toward a dynamical dark energy component or new physics influencing the expansion history at late times.
\item  Utilizing the standard relation between the angular diameter distance $D_A(z)$ and the comoving distance $D_M(z)$, one can arrive at the relation between $D_A''(z)$ and the deceleration Parameter $q(z)$. We compute the second derivative of $D_A(z)$, as
\begin{equation}
D_A''(z) = \frac{(1+z)^2 D_M''(z) - 2(1+z)D_M'(z) + 2D_M(z)}{(1+z)^3} \, ,
\end{equation}
On substituting the derivatives of $D_M(z)$ in terms of the $E(z)$, viz. $D_M'(z) = \frac{1}{E(z)}$ and $D_M''(z) = -\frac{E'(z)}{E^2(z)}$, we can rewrite the expression for $D_A''(z)$. Finally, substituting the relation of $q(z)$ into the expression for $D_A''(z)$, we get
\begin{equation}
D_A''(z) = \frac{1}{(1+z)^2}\left[ 2 D_A(z) - \frac{q(z)+3}{E(z)} \right]
\end{equation}
This relation directly connects the $D_A''(z)$ to the deceleration parameter $q(z)$. Utilizing this relation, we can further identify the particular redshift, say $z_c$, at which $D_A''(z_c) = 0$ satisfies
\begin{equation}
    q(z_c) = 2 \, E(z_c) \, D_A(z_c) - 3 \, .
\end{equation}
This equation identifies the redshift $z_c$ at which the $D_A(z)$ curve has an inflection point. Furthermore, one can compute the relation between $D_A(z)$ and $E(z)$ at transition redshift $z_t$, where $q(z_t)=0$, such that
\begin{equation}
E(z_t) = \frac{3}{(1+z_t)^2 \, D_A''(z_t) - 2 \, D_A(z_t)} \, .    
\end{equation} This dual connection provides an important consistency check between geometry and dynamics in the reconstructed expansion history.
We find $z_t=0.689^{+0.12}_{-0.07}$ with MTGP RQD kernel, and $z_t = 0.723^{+0.066}_{-0.055}$ with knot-based (spline order $k=4$) reconstruction respectively.

\item Figure \ref{fig:rec_set2} illustrates the reconstructed evolution the dark energy density $\rho_{\text{DE}}(z)$, pressure $p_{\text{DE}}(z)$ and energy condition $\rho_{\text{DE}}(z)+p_{\text{DE}}(z)$, using both the MTGP (top row) and knot-based spline (bottom row) approaches. 
\item The middle column of Fig. \ref{fig:rec_set2} illustrates the evolution of the dark energy pressure $p_{\text{DE}}(z)$. This can be reconstructed directly from $E(z)$ and $E(z)^{'}$ without any further inputs. Both the MTGP and knot-based spline reconstructions exhibit a clear deviation in $p(z)$ from the Planck $\Lambda$CDM prediction of constant negative pressure. Specifically, there are deviation from constant pressure $\Lambda$CDM model at more than $2\sigma$ confidence interval for $z<0.25$ as well as around $z \sim 1.0$. This behavior suggests a definite signal for dynamically evolving dark energy.

\begin{figure}[t]
    \centering
    \includegraphics[width=0.325\linewidth]{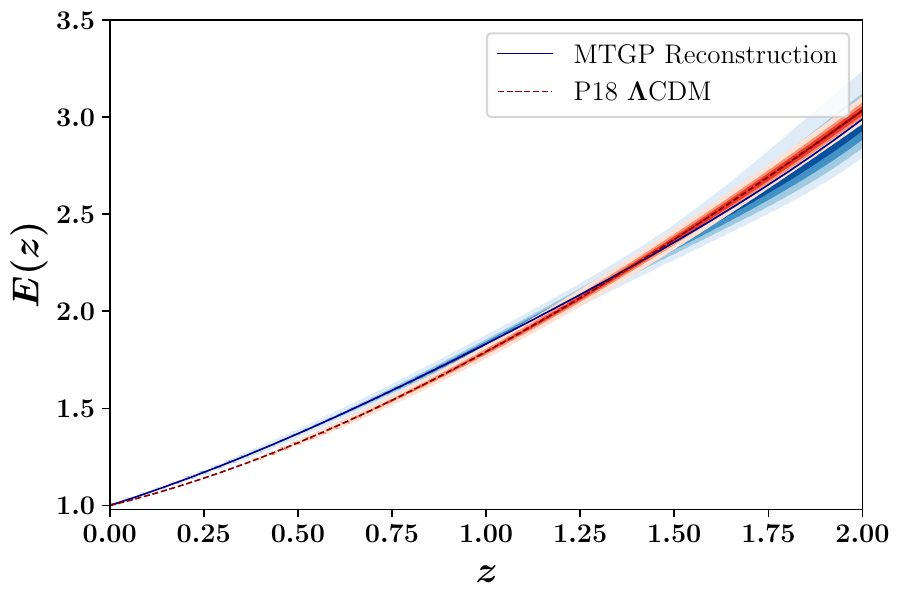}
    \includegraphics[width=0.325\linewidth]{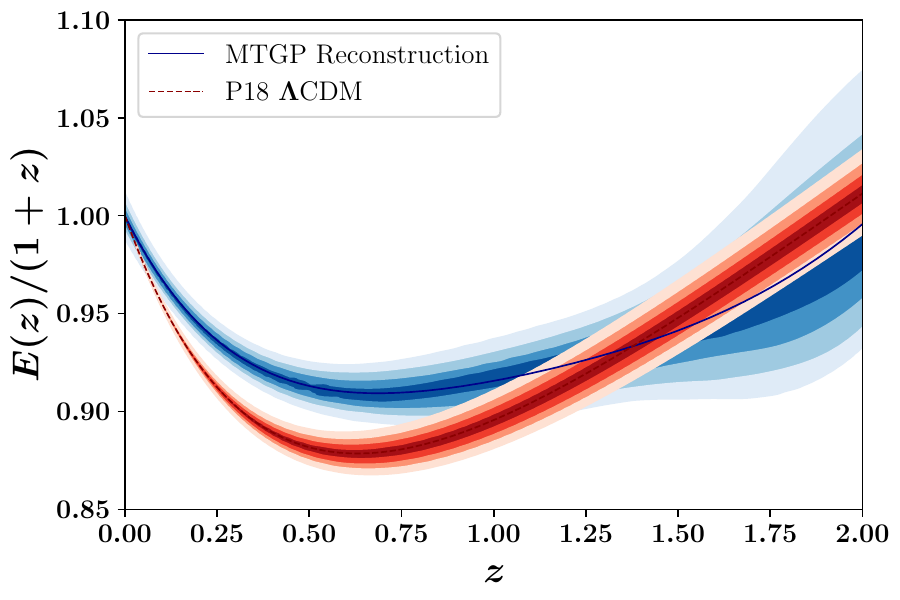}
    \includegraphics[width=0.325\linewidth]{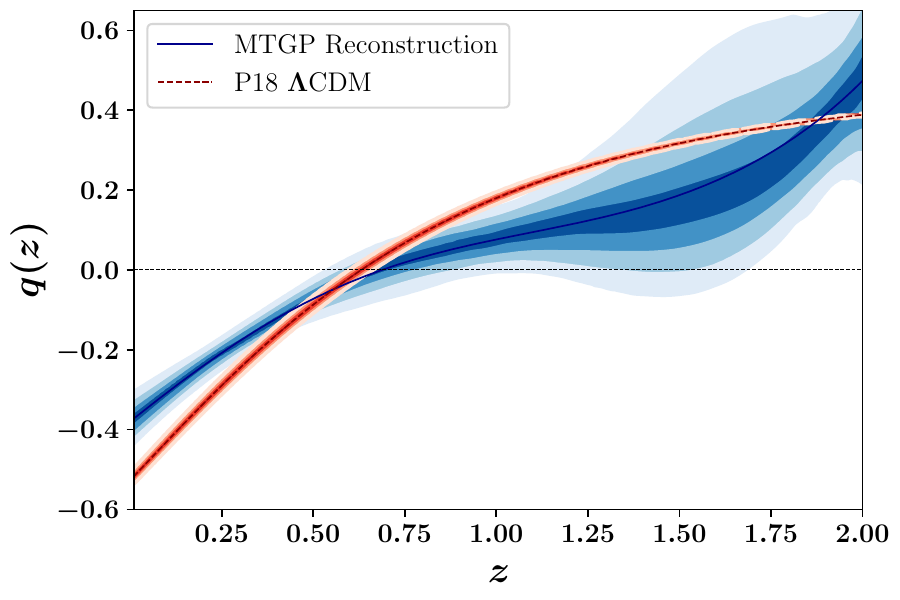} \\
    \includegraphics[width=0.325\linewidth]{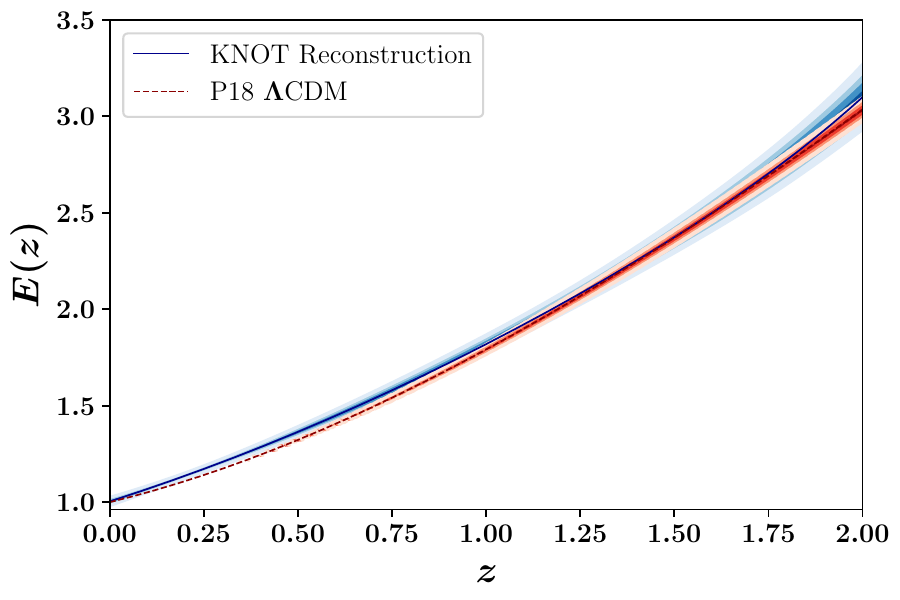}
    \includegraphics[width=0.325\linewidth]{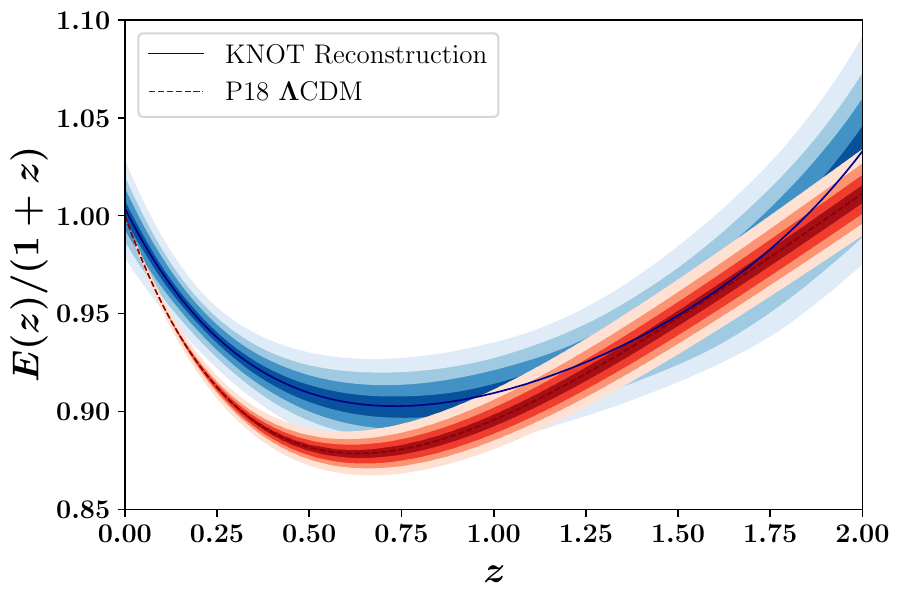}
    \includegraphics[width=0.325\linewidth]{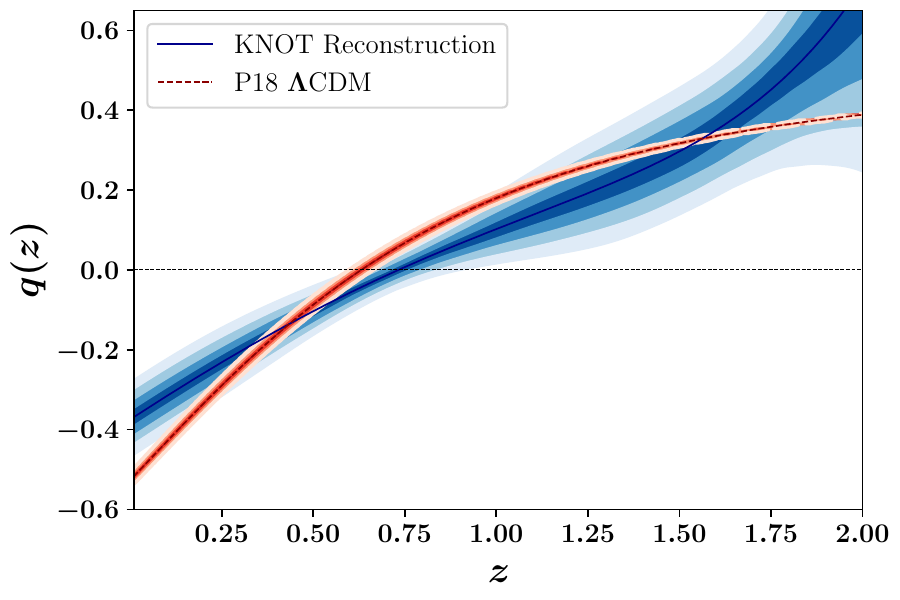}
    \caption{Reconstruction of the reduced Hubble parameter $E(z)$, reduced Hubble parameter $E(z)$ scaled \textit{w.r.t.} $(1+z)$ i.e. $E(z)/(1+z)$, and deceleration parameter $q(z)$, employing (i) MTGP reconstruction with RQD kernel [in upper panel], and (ii) Free-form Knot-based spline reconstruction with $k=4$ order [in lower panel] respectively.}
    \label{fig:rec_set1}
\end{figure}

\begin{figure}[t]
    \centering
    \includegraphics[width=0.325\linewidth]{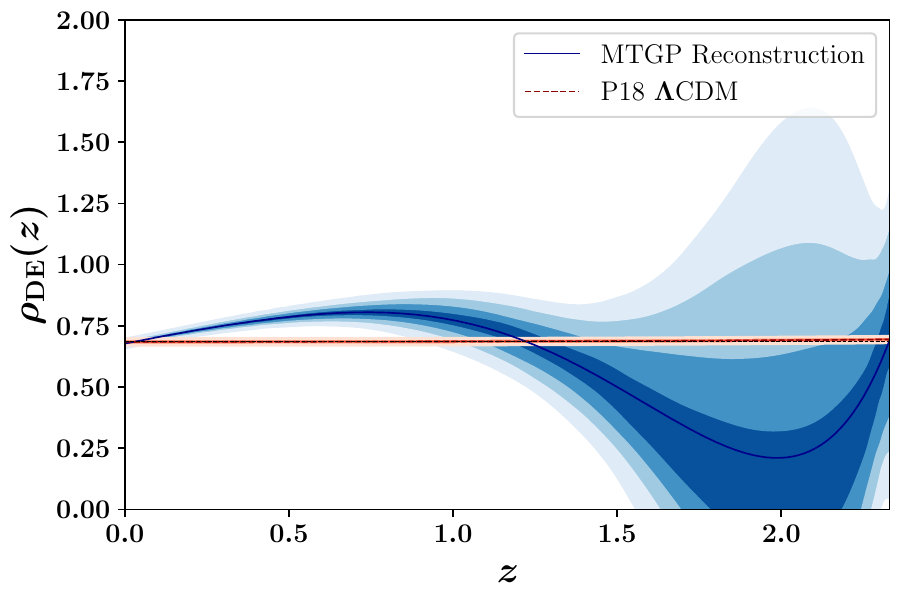}
    \includegraphics[width=0.325\linewidth]{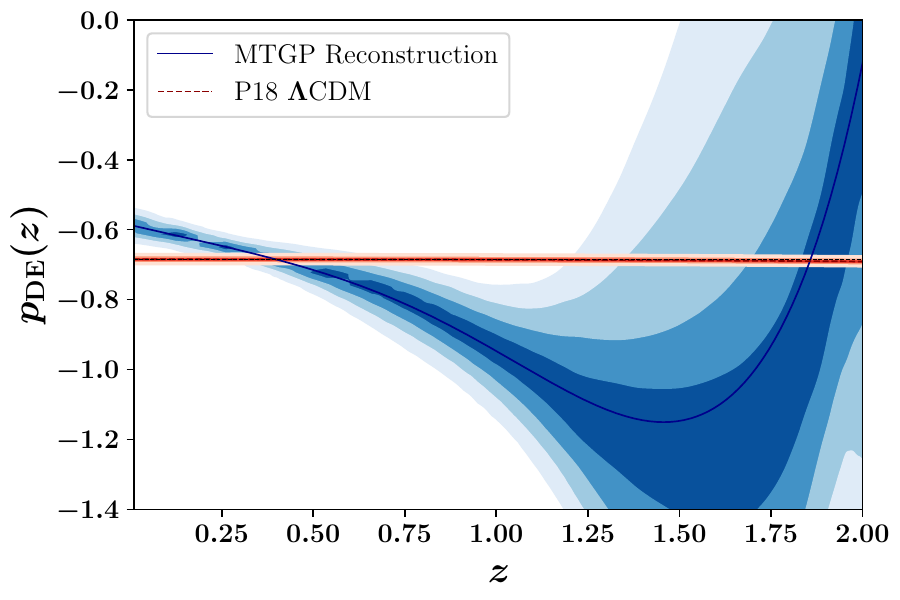} 
    \includegraphics[width=0.325\linewidth]{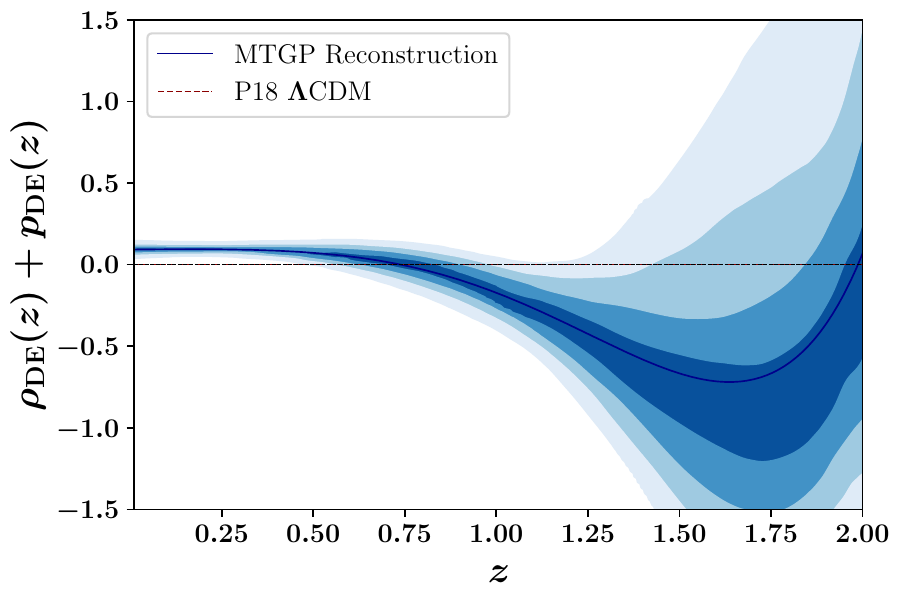} \\
    \includegraphics[width=0.325\linewidth]{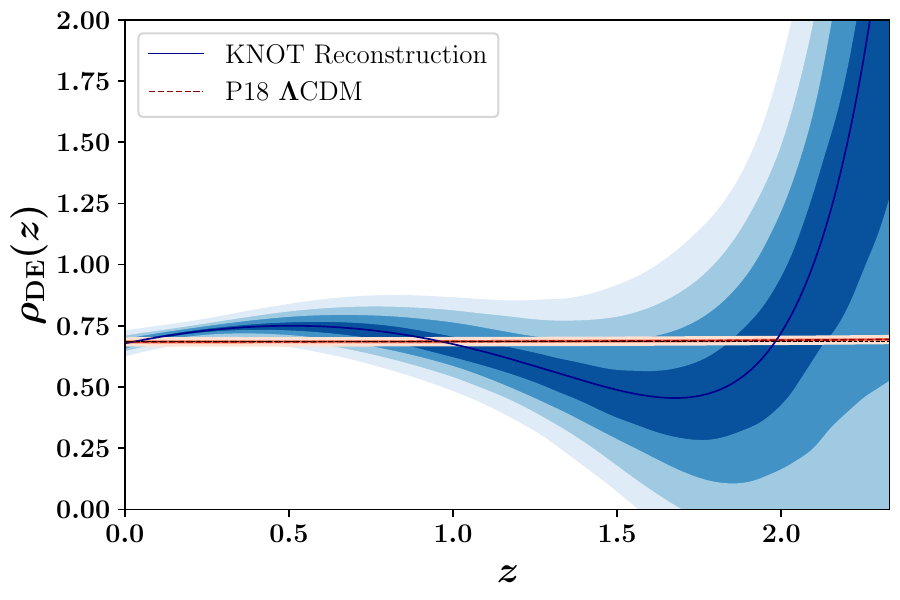}
    \includegraphics[width=0.325\linewidth]{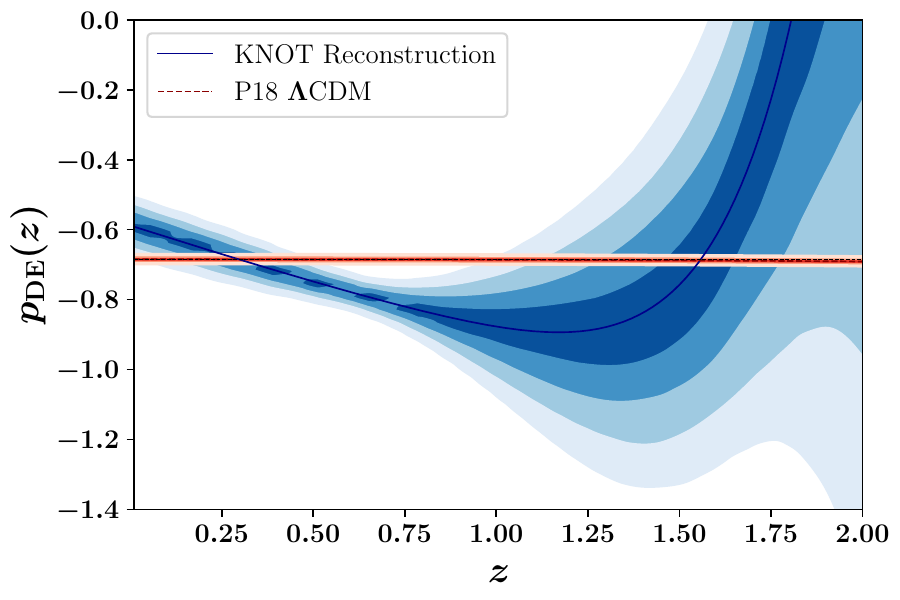} 
    \includegraphics[width=0.325\linewidth]{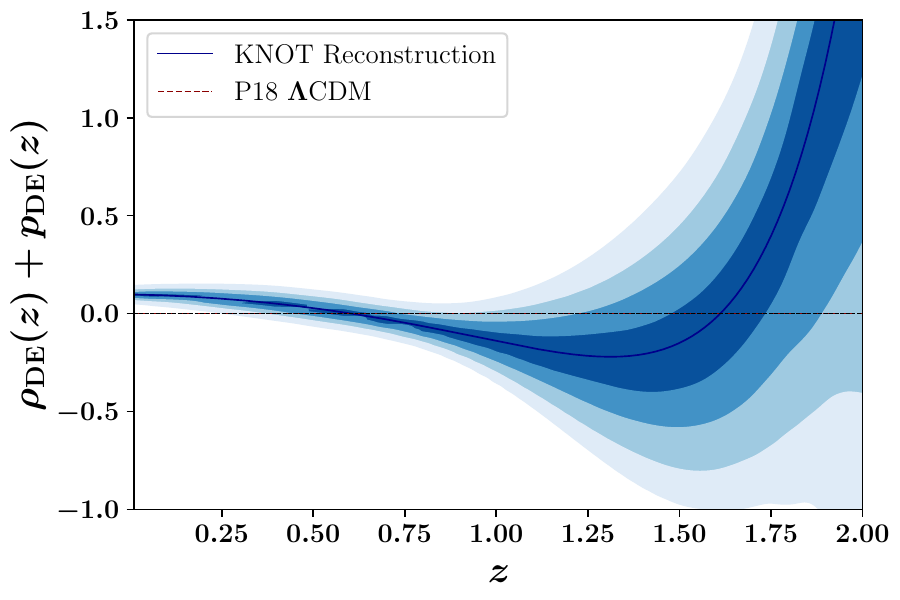} 
    \caption{Reconstruction of dark energy density $\rho_{\rm DE}$, dark energy pressure $p_{\rm DE}$ and evolution of $\rho_{\text{DE}}(z)+p_{\text{DE}}(z)$, employing (i) MTGP reconstruction with RQD kernel [in upper panel], and (ii) Free-form Knot-based spline reconstruction with $k=4$ order [in lower panel] respectively. All quantities are scaled by the critical energy density at present $\rho_{c0} = 3H_{0}^2/8\pi G$.}
    \label{fig:rec_set2}
\end{figure}

\item To extract the dark energy density $\rho_{\text{DE}}(z)$, we require knowledge of the present-day matter density parameter $\Omega_{m0}$. For this, we adopt a Planck $\Lambda$CDM prior on $\Omega_m h^2$, while inferring $H_0$ directly from our reconstructions, assuming that the sound horizon at the drag epoch $r_d$ is calibrated using early-universe physics. This allows us to determine $\Omega_{m0}$, and hence reconstruct $\rho_{\text{DE}}(z)$.
\item The left column of Fig. \ref{fig:rec_set2} presents the reconstructed dark energy density $\rho_{\text{DE}}(z)$. Unlike the constant $\rho_{\text{DE}}$ expected in $\Lambda$CDM, both knot-based spline and MTGP reconstructions show a redshift evolution, the latter significant at $> 2\sigma$ CL.  We observe a gradual increase in the best-fit curve of $\rho_{\text{DE}}$ from $z=0$ up to $z\sim 1$, compared to Planck 2018 $\Lambda$CDM, after which it further decreases, showing a non-monotonic evolution. 

\begin{figure}[t]
    \centering
    \includegraphics[width=0.4\linewidth]{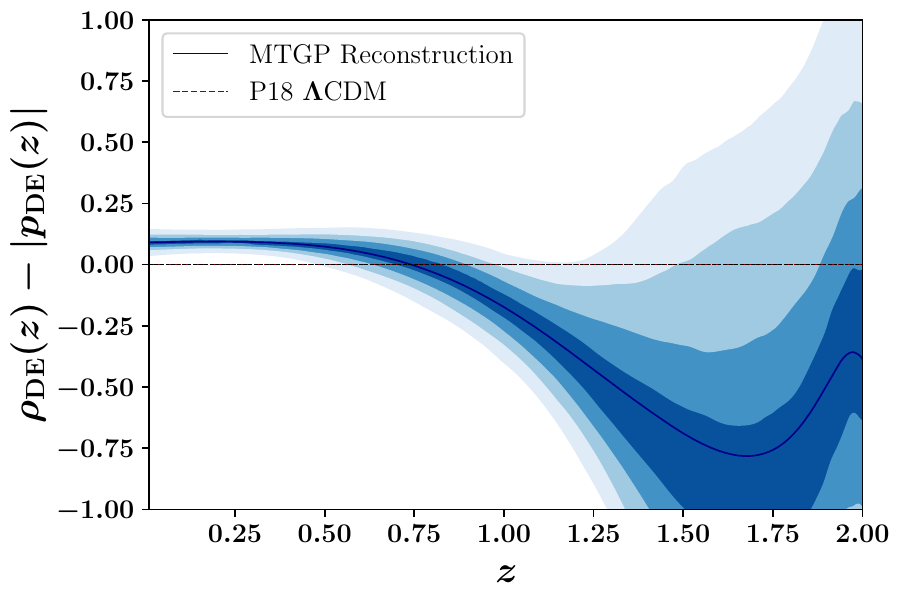}
    \includegraphics[width=0.4\linewidth]{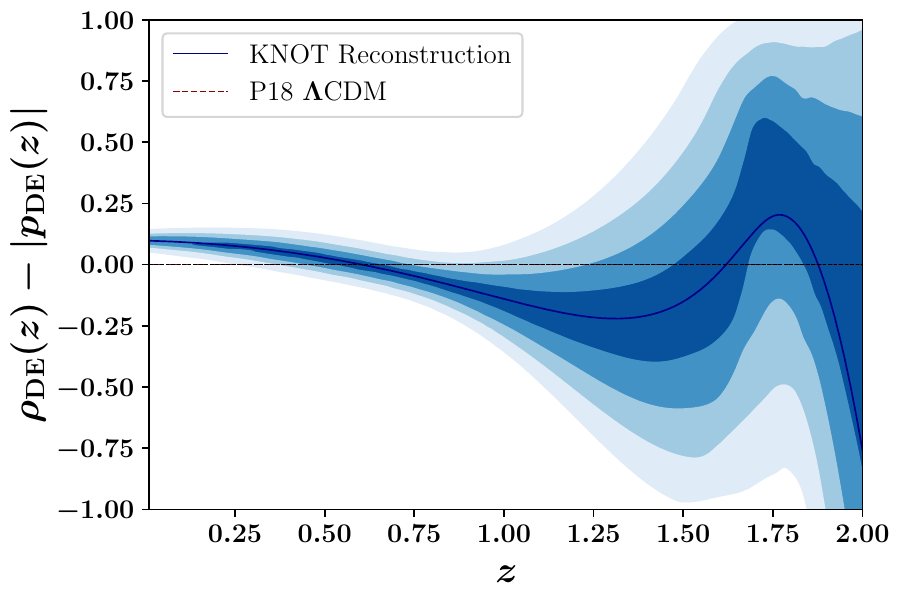} 
    \caption{Reconstruction of quantity  $\rho_{\text{DE}}(z)-\vert p_{\text{DE}}(z)\vert$, the dominant energy condition (DEC) criteria employing (i) MTGP reconstruction with RQD kernel [in left panel], and (ii) Free-form Knot-based spline reconstruction with $k=4$ order [in right panel] respectively.}
    \label{fig:rec_set3}
\end{figure}

\item While $\rho_{\text{DE}}(z)$ does not always show clear tension with $\Lambda$CDM, the reconstructed dark energy pressure $p_{\text{DE}}(z)$ consistently deviates from the Planck $\Lambda$CDM predictions, irrespective of the reconstruction technique employed, indicating a robust tension in the pressure sector. This difference arises because $\rho_{\text{DE}}(z)$ relies on external inputs, such as $\Omega_m h^2$, $H_0$, and the calibrated $r_d$, whose uncertainties propagate into the reconstruction, supposedly masking deviations. In contrast, $p_{\text{DE}}(z)$ is less sensitive to these assumptions, making its deviations more direct and statistically significant.
\item To investigate the dynamical nature of dark energy in a model-independent manner (specifically dark energy exhibit a phantom behaviour), we analyze the redshift evolution of the quantity $\rho_{\text{DE}}(z) + p_{\text{DE}}(z)$, a key diagnostic of the null energy condition (NEC) \cite{Arefeva:2006ido, Qiu:2007fd, Ye:2025ulq}. This combination also serves as a proxy for the product $(1 + w_{\text{DE}})\rho_{\text{DE}}$, allowing us to infer the nature of dark energy—whether it behaves like quintessence ($w_{\text{DE}} > -1$) or exhibits phantom-like features ($w_{\text{DE}} < -1$)—without explicitly reconstructing the equation of state $w_{\text{DE}}$ (Note that $\rho_{DE}$ is always positive as shown in the left panel of Fig.~\ref{fig:rec_set2}.
\item The right panel of Fig.~\ref{fig:rec_set2} shows the reconstructed behavior of $\rho_{\text{DE}} + p_{\text{DE}}$ (to check the validity of the null energy condition (NEC) in the dark energy sector) using both the MTGP and knot-based spline methods. At lower redshifts, ($z < 0.75$), both reconstructions yield $\rho_{\text{DE}} + p_{\text{DE}}$$ \neq 0$ showing a clear deviation from cosmological constant ($\Lambda$) behaviour. Moreover $\rho_{\text{DE}} + p_{\text{DE}} > 0$ for $z \leq 0.75$ confirming the validity of the NEC and hence the non-phantom behaviour for the dark energy at lower redshifts. At higher redshifts, although the violation of the NEC, $\rho_{\text{DE}} + p_{\text{DE}} < 0$ and hence phantom behaviour for dark energy is allowed (as also the phantom crossing),  $\rho_{\text{DE}} + p_{\text{DE}} > 0$ is also allowed within $2\sigma$ confidence interval (especially for the knot based reconstruction). This obviates the need for any unphysical phantom behaviour in the dark energy sector and a minimally coupled canonical scalar field can be a possible candidate for dark energy. This is in contrast to the results obtained using CPL parametrisation where early time phantom behaviour is always needed to fit the DESI-DR2+DESSN-5YR+CMB data \cite{DESI:2025zgx}.
\item Finally in Fig~\ref{fig:rec_set3}, we show the validity of the dominant energy condition: $\rho_{\rm DE} - |p_{\rm DE}| \geq 0$ in the reconstructed dark energy behaviour. It restricts the superluminal propagation of the energy ( in other words, it states that energy will always flow in the time-like or null directions). As shown in Fig~\ref{fig:rec_set3}, although violation of this condition is allowed at high redshifts for dark energy, but within $2\sigma$ confidence level the reconstructed dark energy behaviour also satisfies this condition throughout. Therefore, we do not need any unphysical superluminal propagation of energy in the dark energy sector.
\end{itemize}
To summarise, these results offer multiple, independent hints of potential deviations from the concordance model, particularly at low redshifts. The recurrence of these features across two independent reconstruction methods suggests that these anomalies are not reconstruction artefacts, but may reflect genuine new physics beyond the 6-parameter Planck baseline $\Lambda$CDM model. Moreover the dynamical dark energy behaviour is consistent with all the necessary energy conditions ($\rho_{\rm DE} \geq 0$, $\rho_{\rm DE}+p_{\rm DE} \geq 0$ and $\rho_{\rm DE} \geq |p_{\rm DE}|$) obviating any need for unphysical phantom behaviour or superluminal propagation of energy.

\section{Testing for DESI DR2 BAO Anomalies \label{sec:anomalies}}

Recent studies have raised concerns about possible anomalies in the DESI DR2 BAO measurements at intermediate redshifts, particularly around $z= 0.510$ and $z=0.706$. These redshift bins—corresponding to the LRG1 and LRG2 tracer samples- have been identified as potential outliers in model fits to the full DESI dataset or cross-comparisons with external cosmological probes  \cite{Sapone:2024ltl, Colgain:2025nzf, Colgain:2024mtg, Wang:2024pui, Wang:2024rjd, Colgain:2024xqj}. Such deviations, albeit mild, have drawn attention in the literature due to their potential implications for the expansion history and possible hints of physics beyond $\Lambda$CDM. Specifically, the $d_M/r_d$ values at $z=0.51$ and $d_H/r_d$ values at $z=0.706$ bins appear to lie slightly off the smooth evolution expected from Planck $\Lambda$CDM, raising the question of whether these discrepancies are statistical fluctuations, systematics, or signs of new phenomena.

To assess the validity of these claims, we carry out a model-independent test using the DES-5YR SN-Ia data to reconstruct $D_M(z)$ and $D_H(z)$ via both single-task GP (STGP) regression and a knot-based spline interpolation. The reconstructed values of $D_M(z)$ and $D_H(z)$ (at the LRG1 and LRG2 effective redshift bins) from DES-5YR-SN data is mentioned in Table \ref{tab:desi_anomalies}, with superscripts \textsc{gp} and \textsc{knot}, depending on the reconstruction framework. The reconstructed DES-5YR SN values are then compared directly to the DESI DR2 BAO measurements at $z = 0.510$ and $z = 0.706$. We consider the Planck prior on $h r_d = \mathcal{N}(99.078, \, 0.925)$ \cite{Planck:2018vyg}, where $h = \frac{H_0}{100 \text{ km Mpc$^{-1}$ s$^{-1}$}}$ is the dimensionless Hubble parameter and $r_d$ is the comoving sound horizon at drag epoch, to convert the DESI DR2 BAO $d_M/r_d$ and $d_H/r_d$ measurements to normalized (dimensionless) distance measures $D_M(z)$ and $D_H(z)$. We report these values in Table \ref{tab:desi_anomalies} with superscript \textsc{desi}.

\begin{table}[t]
\centering
\renewcommand{\arraystretch}{1.25}
\setlength{\tabcolsep}{10pt}
\begin{tabular}{c c c c c c c}
\toprule
\textbf{Tracer} & $z_{\rm eff}$ & $D_M^{\rm DESI}$ & $D_M^{\rm STGP \vert\,  DES}$ & \textbf{T ($\sigma$)} & $D_M^{\rm KNOT \vert\,  DES}$ & \textbf{T ($\sigma$)} \\
\midrule
LRG1 & 0.510 & $0.449 \pm 0.007$ & $0.436 \pm 0.003$ & 1.81  &  $0.435 \pm 0.004$ & 1.74 \\
LRG2 & 0.706  & $0.573 \pm 0.007$ & $0.572 \pm 0.004$ & 0.13 & $0.574 \pm 0.005$ & 0.12 \\
\midrule
\textbf{Tracer} & $z_{\rm eff}$ & $D_H^{\rm DESI}$ & $D_H^{\rm STGP \vert\,  DES}$ & \textbf{T ($\sigma$)} & $D_H^{\rm KNOT \vert\,  DES}$ & \textbf{T ($\sigma$)} \\
\midrule
LRG1 & 0.510 & $0.722 \pm 0.015$ & $0.738 \pm 0.009$ & 0.92 & $0.745 \pm 0.014$ & 1.07 \\
LRG2 & 0.706  & $0.643 \pm 0.012$ & $0.640 \pm 0.014$ & 0.16 & $0.667 \pm 0.022$ & 0.96 \\
\bottomrule
\end{tabular}
\caption{Comparison of $D_M$ and $D_H$ values from DESI-DR2 BAO and STGP vs Knot-based reconstructions of DES-5YR SN. Tensions (\textbf{T}) are expressed in units of standard deviation $\sigma$.} \label{tab:desi_anomalies}
\end{table}

The results summarized in Table \ref{tab:desi_anomalies} compare the DESI BAO values of $D_M$ and $D_H$ with our STGP and Knot-based reconstructions with DES SN data. Each entry reports the central value and 1$\sigma$ uncertainty, along with the Gaussian tension metric $\mathrm{\bf T}$ (in units of $\sigma$) between DESI and the reconstructed values from DES. The upper portion of the table presents results for $D_M$, while the lower portion covers $D_H$. The tension metric $\mathrm{\bf T}$ helps in quantifying the level of agreement between the DESI BAO and DES SN-based reconstructions. Our analysis reveals no significant discrepancies: all reconstructed $D_M$ and $D_H$ values agree with the DESI measurements within 2$\sigma$ CL. The largest deviation is observed in $D_M$ at $z = 0.510$, but even this remains well below the threshold of statistical significance, suggesting good consistency between both the datasets --- DESI BAO and DES-5YR SN. These findings suggest that, contrary to earlier speculation, the DESI BAO data at these redshifts do not exhibit anomalous behaviour when evaluated against independent and precise high-quality DES-5YR SN data.

\section{Revisiting the Cosmic Distance-Duality Relation}\label{sec:cddr}

The cosmic distance-duality relation (CDDR), also known as Etherington’s reciprocity theorem \cite{2007GReGr..39.1047E}, is a foundational geometric (theoretical) identity in relativistic cosmology. It relates the luminosity distance $d_L(z)$ and the angular diameter distance $d_A(z)$ via the expression:
\begin{equation}
d_L(z) = (1+z)^2 d_A(z) \, , \label{eq:cddr}
\end{equation}
This relation is a direct consequence of photon number conservation and light propagation along null geodesics in any metric theory of gravity --- particularly within the FLRW framework --- and must hold independently of the specific cosmological model.

However, when combining observational data from different sources, such as SN-Ia (which constrain $d_L$) and BAO (which constrain $d_A$ via $d_M$), it is crucial to test the CDDR explicitly \cite{Bassett:2003vu}. These observables are derived from distinct astrophysical processes and are subject to independent systematics, calibration strategies, and analysis pipelines. Even small violations of the CDDR, whether from new physics or unaccounted-for systematics, could bias any joint reconstruction of cosmological distances \cite{Afroz:2025iwo, Teixeira:2025czm, Keil:2025ysb, Yang:2025qdg, Alfano:2025gie}.

Following this logic, we test the consistency of the CDDR using DESI DR2 BAO $d_A(z)$ and the reconstructed $d_L(z)$ from DES-5YR SN Ia data. To assess this, we define the dimensionless CDDR diagnostic \cite{More:2008uq, Nair:2015jua, Mukherjee:2021kcu, Kumar:2021djt}:
\begin{equation}
\eta(z) \equiv \frac{(1+z)^2 d_A(z)}{d_L(z)} \, , \label{eq:eta}
\end{equation}
which should be identically equal to unity across redshift if the CDDR is satisfied. Significant deviations from $\eta(z) = 1$ would indicate either a breakdown of the underlying assumptions or unresolved systematics in the data.

\begin{figure}[t]
    \centering
    \includegraphics[width=0.45\linewidth]{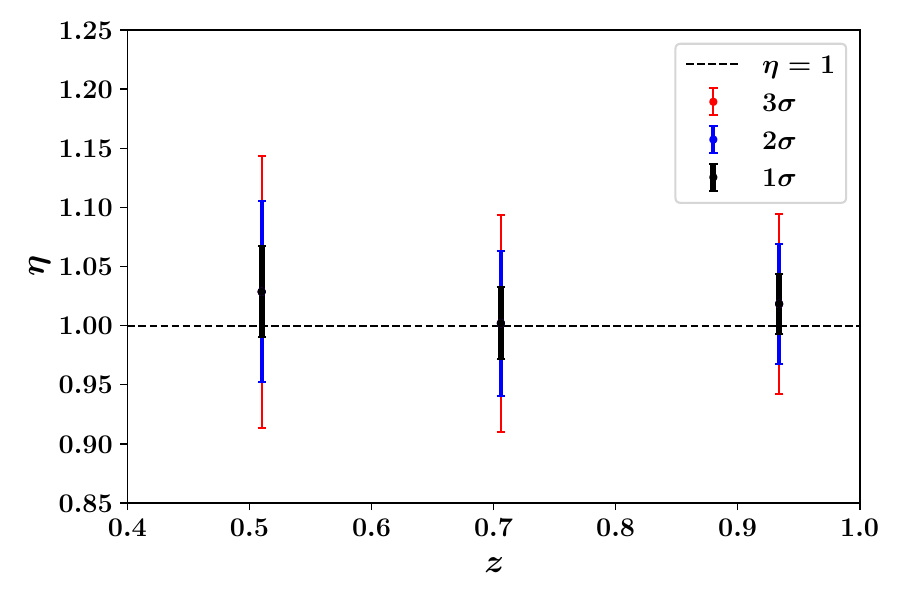}
    \includegraphics[width=0.45\linewidth]{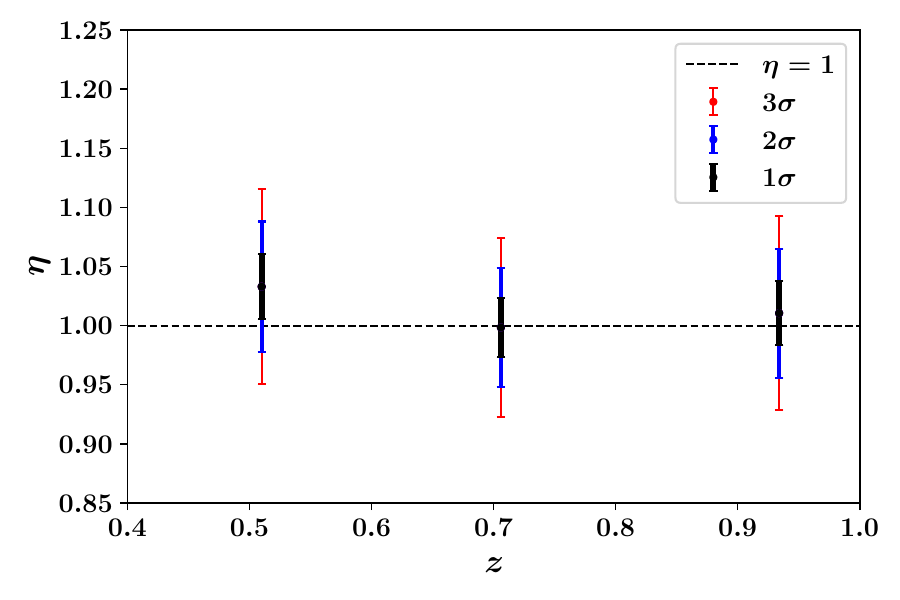}
    \caption{Evolution of $\eta(z)$ across the effective redshift bins ($z_{\rm eff}$) of the DESI-DR2 BAO tracers that overlap with the redshift range covered by the DES-5YR SN-Ia sample, derived from the reconstructed $d_L(z)$ from DES-5YR SN-Ia and $d_A(z)$ from DESI-DR2 BAO measurements, employing STGP (in left panel) and free-form knot reconstruction at characteristic redshift knots.}
    \label{fig:cddr}
\end{figure}

Using both the STGP framework and the knot-based interpolation method to reconstruct $d_L(z)$ from SN Ia data, we compute $\eta(z)$ at the effective redshifts probed by DESI DR2. The uncertainties in $d_A(z)$ and $d_L(z)$, including their covariances, are carefully propagated. The resulting values of $\eta(z)$ are shown in Fig. \ref{fig:cddr}, with the left and right panels corresponding to the GP and knot-based reconstructions, respectively. In both cases, $\eta(z)$ remains consistent with unity within the $2\sigma$ CL across the full redshift range, supporting the validity of the CDDR within current observational precision.

Given this consistency with the CDDR and the lack of significant tension between the SN and BAO data at $z = 0.510$ and $z = 0.706$ (as shown in Sec. \ref{sec:anomalies}), we conclude that the two datasets are both geometrically and statistically compatible. This confirmation retroactively validates the model-independent reconstruction of the comoving distance $d_M(z)$ and its derivative $d_M'(z)$ presented in the preceding sections.

\section{Summary and Discussions \label{sec:summary}}

In this work, we present cosmology-independent reconstruction of the expansion history of the Universe, leveraging the latest data from DESI-DR2 BAO and DES-SN5YR observations. We focus on the determination of characteristic redshifts --- specific redshift points, where observational data allows for high-precision estimates for the expansion rate, solely governed by the underlying FLRW geometry. Our approach combines two complementary reconstruction techniques: GP regression and knot-based spline interpolation. These methods are utilized to infer cosmic distances and their derivatives, allowing for direct estimation of $D_A(z)$ and $E(z)$ at discrete redshift nodes without assuming a background cosmological model. Thus, the precise location of these characteristic redshifts will, in turn, help probe the underlying model of our Universe. 

We identified seven such characteristic redshift points between $z \sim 0.35$ and $z \sim 1.62$, which reveal consistent and statistically significant deviations from the Planck 2018 $\Lambda$CDM predictions in the range $z \sim 0.35–0.55$. At these redshifts, we find tensions in $D_A(z)$ are 3-4$\sigma$, and $E(z)$ reaching 4–5$\sigma$, with the most notable discrepancies at $z_1 \sim 0.35$, where $D_A'(z) \approx D_L(z)$, $z_2 \sim 0.41$, where $D_A'(z) \approx D_M(z)$, and $z_3 \sim 0.51$, where $D_A'(z) \approx D_A(z)$. These deviations persist across both reconstruction methods and are robust to kernel choices and data subsampling, indicating that they are unlikely to be artifacts of statistical noise or methodological bias.

Beyond this low redshift range $z \sim 0.35–0.55$, as we move toward higher redshifts, the tension begins to ease. At $z_4 \sim 0.55$ and $z_5 \sim 0.79$, although we find a $>3\sigma$ deviation in $D_A(z)$, but a milder $\sim 2-2.5\sigma$ deviation in $E(z)$, suggesting a gradual return toward $\Lambda$CDM behavior. At higher redshifts ($z_6 \sim 1.24$ and $z_7 \sim 1.63$), both $D_A(z)$ and $E(z)$ show good agreement with standard model expectations, with tensions well below $2\sigma$.

Our findings point towards a statistically significant deviation in the normalized expansion rate that extends across the range $0.35 \lesssim z \lesssim 0.78$, where the Hubble parameter is larger than the Planck 2018 $\Lambda$CDM predictions. This is consistent with $H_{0}$ tension as predicted by SH0ES observations. Our results therefore confirm a Hubble tension starting around $z \lesssim 0.78$ and extending up to $z=0$. On the other hand, for higher redshifts $z > 0.78$, expansion rate of the Universe is consistent with Planck 2018 $\Lambda$CDM predictions and this is mainly due to larger error bars in $D_{A}(z)$ and $E(z)$ (due to less available data at such redshifts). We need more measurements of background expansion at $z\geq 0.8$ to confirm whether the Hubble tension at low redshifts continues to the higher redshifts.

On the other hand, the reconstructed behaviour for $\rho_{\rm DE} + p_{\rm DE}$ is tightly constrained at low redshifts and is $>0$ confirming the non-phantom behaviour with high significance. Given that, we use the early Universe anchor for acoustic scale as derived using Planck 2018 $\Lambda$CDM predictions, the higher $H(z)$ values at low redshifts need to be compensated by lower $H(z)$ at higher redshifts to keep the acoustic scale anchor satisfied. This appears in the reconstructed $\rho_{\rm DE} + p_{\rm DE}$ behavior around $1 \leq z \leq 2$ where a limited violation of null energy condition (NEC) has appeared (although with larger error bars) showing possible phantom behaviour for a limited period of time before the Universe again enters the non-phantom region at higher redshifts. Hence, our model independent reconstruction shows a limited violation of NEC for a limited period of time around $1 \leq z \leq 2$. This can be important for possible theoretical model building for late time DE behaviour. This feature, particularly stands in contrast to the DESI DR2 results using the CPL parameterization, where the dark energy equation of state is non-phantom at present but transitions to a phantom regime at earlier times---and crucially, remains phantom without reverting to non-phantom behavior at higher redshifts. 

To assess the potential sources of redshift-dependent systematics, we performed robustness checks by either removing specific DESI BAO tracers (e.g., LRG1, LRG2) or applying magnitude offset corrections to low-$z$ DES-SN5YR data; however, the tension around $z \sim 0.35$ and $0.41$ persisted in all cases. We tested for anomalies in the BAO data by comparing the DESI measurements of $D_M$ and $D_H$ against our reconstructions using only DES-SN5YR data. We found no statistically significant discrepancies, with deviations remaining below $2\sigma$ CL. This consistency suggests that the DESI BAO data are not anomalous and are in agreement with the supernova-based reconstruction. We further tested the validity of CDDR by computing $\eta(z) = d_L(z)/\left[(1+z)^2 d_A(z)\right]$ at each DESI redshift using the independently reconstructed $D_L(z)$ and $D_A(z)$ at the BAO tracers' effective redshifts. We find $\eta$ remains consistent with unity within $2\sigma$ CL, supporting the internal consistency of the data and the validity of CDDR.

Although these findings are robust, it is important to consider potential systematic uncertainties inherent to the DESI-DR2 BAO and DES-SN5YR datasets that could influence our results. For the DES-SN5YR sample, systematics such as photometric calibration errors, light-curve modeling uncertainties, host-galaxy correlations, and selection biases may affect the inferred luminosity distance and their redshift dependence. Similarly, DESI DR2 BAO measurements can be impacted by assumptions in the BAO signal reconstruction, modeling of nonlinear structure growth, and redshift-dependent selection effects. While these systematics are carefully characterized and minimized by both survey collaborations, subtle residual biases could affect the precision of the reconstructed expansion history. If not fully accounted for, such effects may contribute to or mimic the observed tensions with Planck $\Lambda$CDM. Ongoing improvements in data calibration, modeling, and cross-survey validation will be essential to further strengthen the reliability of these conclusions.

Therefore, one of the key implications of our findings is the possible need for new late-time physics. Modifications to the early Universe, such as a rescaling of the sound horizon $r_d$, are insufficient to explain the localized nature of the discrepancies around $z \sim 0.35–0.55$. Instead, the evidence favors new low-redshift physics---involving deviations from a cosmological constant to evolving dark energy, presence of non-gravitational interactions, or newer degrees of freedom in the cosmic dark sector.

Finally, the characteristic redshifts identified in our reconstruction serve as sharp diagnostics of possible anomalies in the late-time expansion history. The fully geometric nature of our approach, relying solely on direct observables and their derivatives, minimizes modeling biases and enhances sensitivity to new physics. Whether the observed deviations are early signatures of a breakdown in our concordance framework or traceable to unresolved systematics in both the data sets remains an open question.

It is interesting to note that, in recent studies by Ormondroyd et al.\cite{Ormondroyd:2025exu, Ormondroyd:2025iaf}, the authors performed a non-parametric, free-form reconstruction of the dark energy equation of state. They found an unexpected W-shaped structure in the reconstructed behavior of $w(z)$. It is interesting to note that this W-shaped feature emerges at a redshift around z $\approx$ 0.51, which is similar to our characteristic redshift $z_3 \sim 0.512$, where we observe 4-5$\sigma$ deviation from the Planck $\Lambda$CDM model.

%%%%%%%%%%%%%%%%%%%%%%%%%%%%%%%%%%%%%%%%%%%%%%%%%%%%%%%%%%%%%%
{
Recent studies by \cite{Alfano:2024fzv,Alfano:2024jqn} have reported more conservative constraints on dark energy, despite incorporating DESI (DR1) data. These works cross-correlate with Gamma-Ray Burst (GRB) observations, where GRB luminosity distances are calibrated using Observational Hubble Data (OHD) and DESI-BAO catalogs through B\'{e}zier interpolation techniques. Consequently, the differences in conclusions about the nature of dark energy may stem from the selection and treatment of datasets—particularly the GRBs, which are sensitive to the adopted calibration scheme. In our case, we use SNIa data along with the CMB angular scale as a high-redshift distance anchor, which at this stage is better constraining than GRBs. It would be worthwhile in future work to cross-check our findings with GRB-calibrated constraints to evaluate the influence of such combinations and interpolation choices on the inferred dark energy evolution.

Our findings can be contextualized in light of recent cosmographic analyses such as Ref.~\cite{Luongo:2024fww}, which use Taylor series expansion of cosmological distances to probe the evolutionary history. While their present values of the deceleration parameter $q_0$ and snap $s_0$ are consistent with $\Lambda$CDM, a significant deviation is observed in the present value of jerk $j_0$, raising concerns about internal consistency. Interestingly, our reconstructed $q(z)$ shows $>2\sigma$ deviation from $\Lambda$CDM at $z \lesssim 0.5$, suggesting possible tension not just in $j_0$, but already in $q_0$. These differences may arise from the limitations of Taylor expansions, that are valid only at low redshifts and suffer from convergence and truncation issues. In contrast, our non-parametric reconstruction is valid across the full redshift range and is further anchored at high redshifts using CMB distance priors from Planck 2018 --- the angular acoustic scale at the last scattering surface --- keeping the uncertainties in check, a key advantage over cosmography. A similar reconstruction of $j(z)$ following Ref.~\cite{Mukherjee:2020ytg} could be pursued as future work to clarify the nature of the tension in $j_0$. Our findings suggest that cosmography and direct reconstruction provide complementary perspectives, probing different aspects of the data's information content.
}
%%%%%%%%%%%%%%%%%%%%%%%%%%%%%%%%%%%%%%%%%%%%%%%%%%%%%%%%%%%%%%

\begin{table*}[t]
\centering
\caption{Consistency test for the inference of characteristic redshifts ($z_i$), $D_A(z_i)$ and $E(z_i)$ from only the DESI DR2 BAO observations.}
\renewcommand{\arraystretch}{1.25}
\setlength{\tabcolsep}{6pt}
\begin{tabular}{ccccccc}
\toprule
$z_i$ & $D_A(z_i)$ & $D_A^{\text{P18}}(z_i)$ & Tension ($\sigma$) 
      & $E(z_i)$ & $E^{\text{P18}}(z_i)$ & Tension ($\sigma$) \\
\midrule
$z_1$=$0.347 \pm 0.004$    &  $0.236 \pm 0.001$  & $0.235 \pm 0.002$  &   0.17  &  $1.232 \pm 0.010$  & $1.206 \pm 0.005$  &  2.24  \\
$z_2$=$0.409 \pm 0.006$    &  $0.261 \pm 0.001$  & $0.261 \pm 0.002$ &   0.18  &  $1.284 \pm 0.008$  & $1.252 \pm 0.007$ &  3.01  \\
$z_3$=$0.505 \pm 0.009$    &  $0.292 \pm 0.001$  & $0.294 \pm 0.003$ &  0.63   &  $1.367 \pm 0.006$  & $1.327 \pm 0.010$ &  3.58  \\
$z_4$=$0.540 \pm 0.010$    &  $0.302 \pm 0.001$ &  $0.304 \pm 0.003$  &  0.78  &  $1.397 \pm 0.018$ & $1.355 \pm 0.011$ &  2.05 \\
$z_5$=$0.774 \pm 0.017$    &  $0.350 \pm 0.001$  &  $0.355 \pm 0.003$  &  1.47  &  $1.610 \pm 0.016$  &  $1.564 \pm 0.020$ &  1.77 \\
$z_6$=$1.242 \pm 0.029$    &  $0.392 \pm 0.003$  &  $0.397 \pm 0.002$  & 1.56  &  $2.056 \pm 0.051$  &  $2.059 \pm 0.039$ &  0.05 \\
$z_7$=$1.659 \pm 0.079$    &  $0.399 \pm 0.004$  &  $0.403 \pm 0.002$  & 0.81 &  $2.504 \pm 0.024$  & $2.572 \pm 0.106$ & 0.63 \\
\bottomrule
\end{tabular} \label{tab:char_desi}
\end{table*}

Thus, our results add to growing indications of possible inconsistencies between late-time cosmic evolution and Planck $\Lambda$CDM predictions. The synergy between DESI DR2 BAO and DES-SN5YR datasets, and the precision of our geometric reconstructions, enables a robust, model-independent probe of the expansion history --- allowing us to identify subtle, redshift-dependent deviations that may otherwise remain hidden within individual datasets.  Looking ahead, future surveys such as DESI-5YR BAO \cite{Moon:2023jgl}, Euclid \cite{Euclid:2025sit}, Rubin LSST \cite{LSSTDarkEnergyScience:2018jkl, LSSTDarkEnergyScience:2021laz} and 21 cm cosmological inference with SKA \cite{Berti:2022ilk} will dramatically improve statistical precision and redshift coverage, enabling us to confirm, refine, or refute the anomalies reported here. These next-generation datasets will be crucial in disentangling the origin of the observed tensions and in deepening our understanding of cosmic acceleration and the nature of dark energy. \\ 

\begin{figure}[t]
    \centering
    \includegraphics[width=0.95\linewidth]{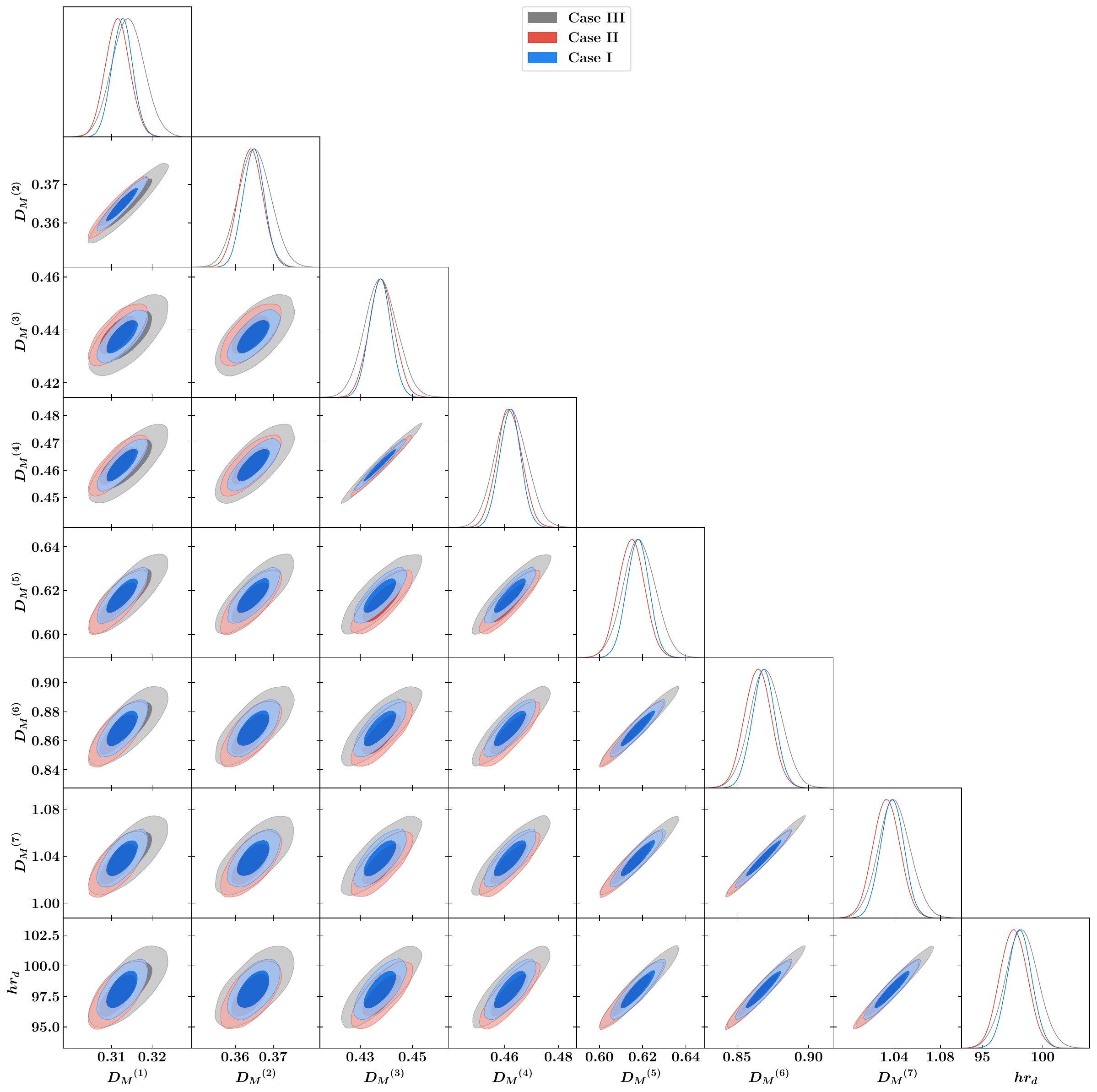} 
    \caption{Joint posterior distributions for $D_M^{(i)}$, $i = 1, \ldots, 7$, at the characteristic redshift knots $z_i$, reconstructed using a knot-based spline method. We compare three cases: I (full data), II (70\% randomly selected DES-SN5YR data), and III (40\% randomly selected DES-SN5YR data), all combined with DESI DR2 BAO measurements. The strong overlap in posteriors indicates statistical consistency among the cases.}
    \label{fig:mcmc_cv_test}
\end{figure}

\begin{figure}[t]
    \centering
    \includegraphics[width=\linewidth]{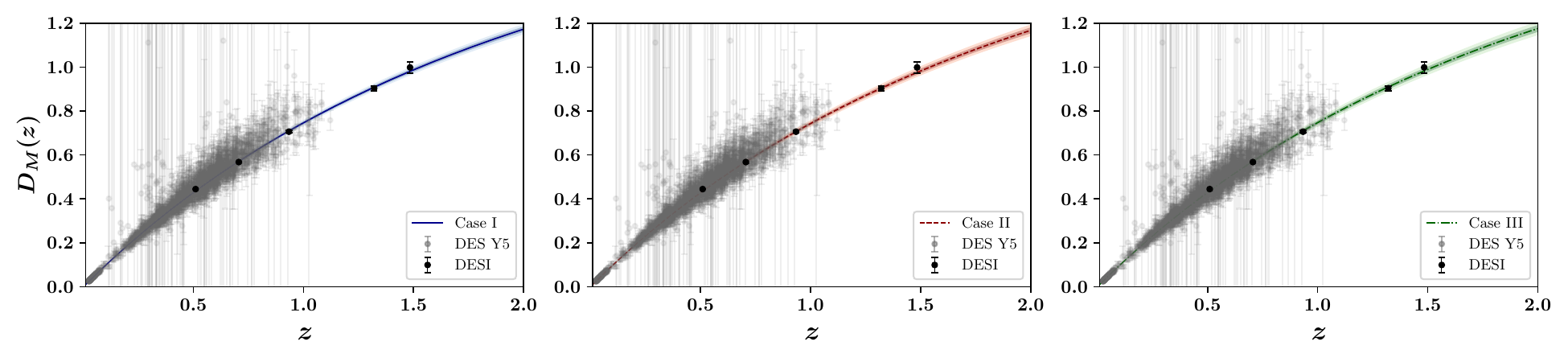}\\
    \includegraphics[width=\linewidth]{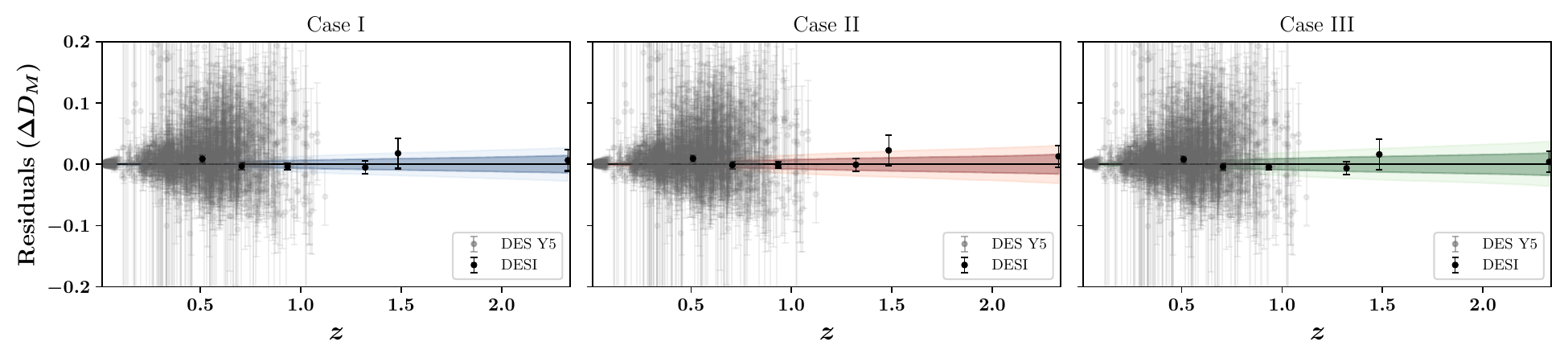}
    \caption{Reconstructed evolution of the dimensionless comoving angular diameter distance $D_M(z)$ for the three data subsets: I (full), II (70\%), and III (40\%) in the top panel. Residuals showing the uncertainty bands of the reconstructions with respect to the data in bottom panel. The reconstructions show excellent agreement, confirming the robustness of the method under data resampling.}
    \label{fig:dm_cv_test}
\end{figure}

\noindent \textbf{Note added:} A recent work by Chavan, Sarkar and Sen \cite{Chavan:2025tis}, uses a semi-cosmographic analysis with Pad\'{e} expansion to reconstruct $D_A(z)$, its derivative $D_A'(z)$, and $f\sigma_8(z)$ from SDSS-IV BAO+RSD data. They identify critical redshifts $z \sim 0.511 \pm 0.052$ where $D_A = D_A'$ and $z \sim 1.623 \pm 0.019$ where $D_A$ is maximum i.e., $D_A’ = 0$, which align closely with our $z_3$ and $z_7$, respectively. While their best-fit values agree with ours, tensions with Planck $\Lambda$CDM are milder due to larger uncertainties in SDSS-IV data compared to DESI DR2. Notably, the tension is significant ($>2\sigma$) at $z\sim 0.51$ but negligible at $z\sim 1.623$, consistent with our findings, given the sensitivity of the data. The agreement across distinct methods and datasets strengthens the case for possible new physics in $z \sim 0.35$–0.55 range and highlights these key redshifts for targeted future observations.

\section*{Data availability statement}
Authors can confirm that all relevant source data from the public domain are included in the manuscript, and/or duly cited.

\begin{acknowledgments}
The authors thank the anonymous reviewers for their insightful comments and constructive suggestions towards the improvement of the manuscript. We acknowledge comments and suggestions from Eoin O Colg\'ain, Eleonora Di Valentino, Shahin Sheikh-Jabbari, Sunny Vagnozzi and Savvas Nesseris. PM acknowledges funding from the Anusandhan National Research Foundation (ANRF), Govt of India, under the National Post-Doctoral Fellowship (File no. PDF/2023/001986). AAS acknowledges the funding from ANRF, Govt of India, under the research grant no. CRG/2023/003984. We acknowledge the use of the HPC facility, Pegasus, at IUCAA, Pune, India. We also acknowledge the CERN TH Department for hospitality and computing support during the later stages of this work. This article/publication is based upon work from COST Action CA21136- ``Addressing observational tensions in cosmology with systematics and fundamental physics (CosmoVerse)'', supported by COST (European Cooperation in Science and Technology).
\end{acknowledgments}

\appendix

\section{Data Subset Consistency and Robustness Check}

In this section, we examine the consistency of our findings using only the DESI DR2 observations. We focus on this particular dataset because it provides combined measurements of $d_M/r_d$, $d_H/r_d$ and their joint covariance. We repeat the entire analysis to infer the characteristic redshifts $z_i$’s and compute the normalized angular diameter distance $D_A(z_i)$ and the reduced Hubble parameter $E(z_i)$. For demonstration purposes, we show the tensions observed across different redshift bins, employing the knot-based spline reconstruction in Table \ref{tab:char_desi}.

Notably, the tension exceeds the 3$\sigma$ CL in the lower redshift range (at $z_1$,$ z_2$, $z_3$). This is consistent with our previous finding using combined DESI DR2 and DES-SN5YR data. With consistency tests of DESI DR2 BAO anomalies and the validity of the Cosmic Distance Duality Relation (CDDR) when comparing DESI and DES observations,  we have shown that two datasets are compatible, showing no significant tension between them, which supports the robustness of our results.

Furthermore, these trends in the expansion rate anomalies at characteristic redshifts $z_1$, $z_2$ and $z_3$ (observed with DESI DR2 BAO) become even more pronounced and clearer with higher-quality DES-SN5YR observations at low redshifts, leading to a significant 4-5$\sigma$ tension with the Planck $\Lambda$CDM model specifically at $z_1$, $z_2$, and  $z_3$, respectively.

\section{Cross-Validation of the Reconstructed Uncertainties}

To evaluate the reliability of the predicted uncertainty bands in the reconstruction of the comoving angular diameter distance, $D_M(z)$, we perform a cross-validation analysis using different subsets of the data. Specifically, we consider the following three cases: 
\begin{itemize}
    \item[\textbf{I.}] Full DES-SN5YR dataset,
    \item[\textbf{II.}] A randomly selected 70\% subset of the DES-SN5YR data,
    \item[\textbf{III.}] A randomly selected 40\% subset of the DES-SN5YR data.
\end{itemize}
In each case, we combine the respective supernova subset with DESI-DR2 BAO data and carry out the full reconstruction of $D_M(z)$ using the knot-based spline method.

Figure~\ref{fig:mcmc_cv_test} shows the joint posterior distributions for the comoving distances $D_M^{(i)}$ at the seven redshift knots $z_i$, comparing the three data subsets. The posteriors show strong overlap, indicating statistical consistency between the cases. In Figure~\ref{fig:dm_cv_test}, we display the reconstructed evolution of $D_M(z)$ as a function of redshift (top panel), along with the residuals quantifying the uncertainty bands relative to the datasets (bottom panel). The reconstructions are in excellent agreement, highlighting the robustness of the method under random data reduction.

%\nocite{*}

\bibliography{refs}

\end{document}